\documentclass[10pt, conference, letterpaper]{IEEEtran}
\usepackage[utf8]{inputenc}
\usepackage[english]{babel}
\usepackage[usenames,dvipsnames]{xcolor}
\usepackage{booktabs}

\usepackage[utf8]{inputenc}
\usepackage{xspace}
\usepackage{subfigure}
\usepackage{graphicx, cite}
\usepackage{balance}
\usepackage{comment}
\usepackage{alltt}
\usepackage{multirow}
\usepackage{CJK}
\usepackage{textcomp}
\usepackage{amssymb}
\usepackage{amsmath}
\usepackage{amsfonts}

\usepackage{fancyhdr}
\usepackage{lastpage}
\newcommand{\BK}[1]{}
\IEEEoverridecommandlockouts
\begin{document}

\title{Architectural Design Alternatives based on Cloud/Edge/Fog Computing for Connected Vehicles}

\author{Haoxin Wang\IEEEauthorrefmark{1}, Tingting Liu\IEEEauthorrefmark{2}, BaekGyu Kim\IEEEauthorrefmark{3}, Chung-Wei Lin\IEEEauthorrefmark{4}, Shinichi Shiraishi\IEEEauthorrefmark{5},\\ Jiang Xie\IEEEauthorrefmark{1}, and Zhu Han\IEEEauthorrefmark{6}\IEEEauthorrefmark{7}\\
\IEEEauthorblockA{\IEEEauthorrefmark{1}University of North Carolina at Charlotte, Charlotte, NC 28223, U.S.A.}
\IEEEauthorblockA{\IEEEauthorrefmark{2}Nanjing Institute of Technology, Nanjing 211167, CHINA}
\IEEEauthorblockA{\IEEEauthorrefmark{3}Toyota Motor North America (TMNA) R\&D InfoTech Labs, U.S.A.}
\IEEEauthorblockA{\IEEEauthorrefmark{4}National Taiwan University, Taipei, Taiwan}
\IEEEauthorblockA{\IEEEauthorrefmark{5}Toyota Research Institute - Advanced Development}
\IEEEauthorblockA{\IEEEauthorrefmark{6}University of Houston, Houston, TX 77004, U.S.A.}
\IEEEauthorblockA{\IEEEauthorrefmark{7}Department of Computer Science and Engineering, Kyung Hee University, Seoul, South Korea, 446-701\\E-mail:~hwang50@uncc.edu;~liutt@njit.edu.cn;~baekgyu.kim@toyota.com;~cwlin@csie.ntu.edu.tw;\\~shinichi.shiraishi@tri-ad.global;~linda.xie@uncc.edu;~hanzhu22@gmail.com}

\thanks{This is a personal copy of the authors. Not for redistribution. The final version of this paper was accepted by IEEE Communications Surveys \& Tutorials and is available through the IEEE Xplore Digital Library, at the link: https://ieeexplore.ieee.org/document/9184917, with the DOI: 10.1109/COMST.2020.3020854.}
}

\maketitle

\thispagestyle{fancy}
\begin{abstract}
    As vehicles playing an increasingly important role in people's daily life, requirements on safer and more comfortable driving experience have arisen. Connected vehicles (CVs) can provide enabling technologies to realize these requirements and have attracted widespread attentions from both academia and industry. These requirements ask for a well-designed computing architecture to support the Quality-of-Service (QoS) of CV applications. Computation offloading techniques, such as cloud, edge, and fog computing, can help CVs process computation-intensive and large-scale computing tasks. Additionally, different cloud/edge/fog computing architectures are suitable for supporting different types of CV applications with highly different QoS requirements, which demonstrates the importance of the computing architecture design. However, most of the existing surveys on cloud/edge/fog computing for CVs overlook the computing architecture design, where they (i) only focus on one specific computing architecture and (ii) lack discussions on benefits, research challenges, and system requirements of different architectural alternatives. In this paper, we provide a comprehensive survey on different architectural design alternatives based on cloud/edge/fog computing for CVs. The contributions of this paper are: (i) providing a comprehensive literature survey on existing proposed architectural design alternatives based on cloud/edge/fog computing for CVs, (ii) proposing a new classification of computing architectures based on cloud/edge/fog computing for CVs: computation-aided and computation-enabled architectures, (iii) presenting a holistic comparison among different cloud/edge/fog computing architectures for CVs based on functional requirements of CV systems, including advantages, disadvantages, and research challenges, (iv) presenting a holistic overview on the design of CV systems from both academia and industry perspectives, including activities in industry, functional requirements, service requirements, and design considerations, and (v) proposing several open research issues of designing cloud/edge/fog computing architectures for CVs.
    
\end{abstract}

\section{Introduction}\label{sec:Introduction}
As vehicles playing an increasingly important role in people's daily life, more requirements, such as higher efficient traffic, safer road, and more comfortable driving experience, have arisen. These requirements may consume a large amount of computation and communication resources. Connected vehicles (CVs), which provide enabling technologies to realize the aforementioned requirements in vehicular networks, have attracted widespread attentions from both academia and industry.

\textit{CVs} are network attached vehicles that exchange data with the cloud and other network attached devices and servers \cite{AECC_2018_WG2}. CVs use different communication technologies to communicate with the driver, other cars on the road. Vehicle-to-everything (V2X) communications, as shown in Table \ref{Table1}, include vehicle-to-vehicle (V2V), vehicle-to-infrastructure (V2I), vehicle-to-cloud (V2C), and vehicle-to-driver (V2D) communications. Traditionally, the automotive industry has been mainly driven by automotive original equipment manufacturers (OEMs) that were capable of producing and maintaining a massive amount of car hardware. However, the trend of CVs will accommodate other types of key players to make it realized, such as governments building roadside infrastructures, telecommunication companies maintaining nation-wide communication infrastructures, and information technology (IT) companies providing various software-based services using a large amount of data. Hence, it is necessary to develop key technologies required to drive the trend in a larger context with such new stakeholders.

\begin{table*}[tp!]
\newcommand{\tabincell}[2]{\begin{tabular}{@{}#1@{}}#2\end{tabular}}
\centering
\caption{Summary of V2X}
\begin{tabular}{|c|c|c|c|c|}
\hline

\multicolumn{2}{|c|}{}& \textbf{Characteristics}  & {\textbf{Research Topics}} & {\textbf{Example Messages}}\\\hline

\multirow{14}{*}{\tabincell{c}{V2X}} & \multirow{5}{*}{\tabincell{c}{V2V}} & \multirow{3}{*}{\tabincell{c}{1. V2V channel is distinct from the typical cellular channel\\ 2. Antenna heights of both transmitter and receiver are low\\ 3. Both transmitter and receiver are mobile\\ 4. Channel characteristics are influenced by different traffic\\ environments \cite{sun2013parking,matolak2008channel}}} & \tabincell{c}{Channel modelling (e.g., expressway,\\ intersection, suburban street, etc.) \cite{sun2013parking,matolak2008channel,acosta2007six, sen2008vehicle,karedal2009geometry,molisch2009survey,mecklenbrauker2011vehicular,karedal2011path}} &  \multirow{5}{*}{\tabincell{c}{Cooperative awareness\\ message \cite{etsi2011intelligent}, Broadcast\\ safety message \cite{chevreuil2002ivhw,basheer2017review},\\ Decentralized environment\\ notification message \cite{etsi2014302}}}\\\cline{4-4}
&   &   & \multirow{2}{*}{Security and privacy \cite{basheer2017review,whyte2013security,rao2007secure,harding2014vehicle,hasrouny2015group,iyer2008secure}}& \\
&   &   & & \\\cline{4-4}
&   &   & \multirow{2}{*}{Testbed \& simulation framework \cite{mir2017large,wang2012improving,kwon2016integrated}} & \\
&   &   & & \\\cline{2-5}

& \multirow{4}{*}{\tabincell{c}{V2I}} & \multirow{4}{*}{\tabincell{c}{1. Data transmission is influenced by Doppler effect \cite{lorca2017overcoming}\\ 2. Require a large uplink capacity \cite{8761551}}} & \multirow{2}{*}{\tabincell{c}{Channel modelling \cite{zhang2016channel,lee2012simulation,viriyasitavat2015vehicular,khabbaz2014modeling}}} &  \multirow{3}{*}{\tabincell{c}{Cooperative traffic safety\\ message \cite{toulminet2008comparative},\\ Edge-assisted service\\ message \cite{Q_2018_IN_Toward,qiu2019secure}}}\\
&   &   &   & \\\cline{4-4}
&   &   & Security and privacy \cite{qiu2019secure,kim2010efficient,ayoub2018security} & \\\cline{4-4}
&   &   & Testbed \& simulation framework \cite{lee2012simulation} & \\\cline{2-5}

& \multirow{4}{*}{\tabincell{c}{V2C}} & \multirow{4}{*}{\tabincell{c}{1. High latency\\ 2. Require routing protocols}} & \multirow{2}{*}{\tabincell{c}{Scheduling \cite{quan2016icn,ibarz2020optimizing}}} &  \multirow{4}{*}{\tabincell{c}{Cloud-assisted service\\ message \cite{wang2020digital}}}\\
&   &   &   & \\\cline{4-4}
&   &   & Security and privacy \cite{rangarajan2012v2c} & \\\cline{4-4}
&   &   & Testbed \& simulation framework \cite{kim2017isv2c} & \\\cline{2-5}

&{\tabincell{c}{V2D}} & {\tabincell{c}{Body area sensor network \cite{hanson2009body}}} & {Testbed \& simulation framework \cite{spelta2010smartphone}} &  \tabincell{c}{Driver health\\ message}\\\hline

\end{tabular}
\label{Table1}
\end{table*}

Therefore, the development of CVs is largely dependent on the information and communication technologies which have fueled a plethora of innovations in various areas, including computing, communication, and caching. Due to the limited on-board battery and computation capacity in vehicles, in order to execute a large number of computations in limited time, offloading power-intensive time-consuming computation tasks to other more powerful servers may significantly improve the performance of many applications of CVs, such as intelligent driving, cruise assist, and high-resolution map creation. Therefore, cloud computing, edge computing, and fog computing are proposed to realize such computation offloading.

\begin{table*}[tp!]
\newcommand{\tabincell}[2]{\begin{tabular}{@{}#1@{}}#2\end{tabular}}
\centering
\caption{The Comparison of Features Among Different Computing Paradigms}
\begin{tabular}{|c|c|c|c|c|}
\hline
\textbf{\multirow{2}{*}{Features}} & \multirow{2}{*}{MCC} & \multicolumn{2}{c|}{Edge Computing} & \multirow{2}{*}{Fog Computing}\\\cline{3-4}
    &     & MEC & Cloudlet &  \\\hline
\textbf{Firstly Proposed By} & Not specific & ETSI & Carnegie Mellon University & Cisco \\\hline 
\textbf{Architecture} & CVs - cloud & CVs - edge servers & CVs - cloudlet nodes - cloud & CVs - fog nodes - cloud\\\hline
{\textbf{\tabincell{c}{Location of\\ Computation Resources}}} & \tabincell{c}{Deployed in a\\ remote data center} & \tabincell{c}{Deployed at\\network edges} &  \tabincell{c}{Attached with APs/BSs} & \tabincell{c}{Near CVs\\(e.g., RSUs, buses, CVs)}\\\hline
\textbf{Operation Mode} & Standalone & Standalone & Standalone/cooperate with cloud & Cooperate with cloud\\\hline 
\textbf{Distance to CVs} & Far & Close & Very close & Very close \\\hline 
\textbf{Service Coverage} & Global & Less global & Local & Local \\\hline 
\textbf{Communication Latency} & High & Low & Low & Low \\\hline 
\textbf{Virtualization Technology} & VM \& others & VM \& others & Only VM & VM \& others \\\hline
\textbf{Main Driver} & Academia & \tabincell{c}{Industry\\(ETSI \cite{ETSI_MEC})} & \tabincell{c}{Academia \& industry\\(Open Edge Computing Initiative \cite{OpenEdge})} & \tabincell{c}{Academia \& industry\\(OpenFog Consortium \cite{OpenFog})} \\\hline
\textbf{\tabincell{c}{Main Issue in\\ CV Scenario}} & \tabincell{c}{Long response latency} &  \multicolumn{3}{c|}{1. Poor mobility support 2. Limited computation \& storage capabilities} \\\hline
\textbf{\tabincell{c}{Main Benefits in\\ CV Scenario}}  & \tabincell{c}{Ample storage \& \\computation capabilities} &  \multicolumn{3}{c|}{\tabincell{c}{1. Real-time data processing \& low-latency response to CVs\\2. Local area information collection, filtering, and cleansing}} \\\hline

\end{tabular}
\label{Table2}
\end{table*}

\emph{Mobile cloud computing} (MCC) which can perform large-scale and computation-intensive computing has been developed over the past decade. Cloud computing is defined as ``a model for allowing ubiquitous, convenient, and on-demand network access to a number of configured computing resources (e.g., networks, server, storage, application, and services) that can be rapidly provisioned and released with minimal management effort or service provider interaction" \cite{mell2009nist}. MCC offers a lot of attractive features, such as parallel processing, virtualized resources, high scalability, and security. Therefore, MCC can not only provide the ability of processing computation-intensive tasks, but also offers low cost infrastructure maintenance \cite{lu2013improvement}. However, nowadays it is predicted that CVs may produce a large amount of data in high speeds such as the camera captured videos for driving assistance, which will make the data dramatically increase to TB/PB levels in seconds \cite{Darwish_2018_Access}. Additionally, a large number of applications of CVs tend to be latency-sensitive and have fast big-data processing with quick response demands. In an intelligent driving scenario, for example, sensors and $3$-D cameras attached to a CV can generate considerably massive data. Thus, the cloud server must complete computing these data and send back highly accurate operating instructions to the CV's steering system in milliseconds level. However, since in terms of network topology, cloud servers are far away from the CVs, a long latency may be caused by the network congestion or queuing, which may, in the worst case, incurs car accidents.

\emph{Multi-access edge computing} (MEC) is an efficient solution to address the aforementioned issues, where a lower response delay can be obtained due to the computation is performed close to CVs, instead of being sent to the remote cloud. The concept of MEC is firstly introduced by the European Telecommunications Standards Institute (ETSI) in $2014$ under the name of Mobile Edge Computing (MEC), where an IT service environment and cloud-computing capabilities can be acquired at the mobile network edges (e.g., the edge of the cellular network) \cite{ESTI_2018_MEC,hu2015mobile}. In $2017$, ETSI officially renamed Mobile Edge Computing to Multi-Access Edge Computing ``to embrace the challenges in the second phase of work and better reflect non-cellular operators’ requirements'' \cite{ESTI_2017_MEC}. Thus, the access approaches in MEC become more variant in CV scenarios. For example, a CV can directly offload its collected raw data to a powerful computing unit that is deployed in a nearby small cell base station (BS) or a roadside unit (RSU). In addition, \emph{Cloudlet} is one of the most typical edge computing platforms, where a cluster of resource-rich computing nodes are placed only one wireless hop away from mobile users (MUs). The computing nodes run multiple virtual machines (VMs) to provide computing services for MUs. ``Essentially, a cloudlet resembles a data center in a box: it is self-managing, requiring little more than power, Internet connectivity, and access control for setup. Internally, a cloudlet resembles a cluster of multicore computers, with gigabit internal connectivity and a high-bandwidth wireless local area network (WLAN)" \cite{Satyanarayanan_2009_PC}. Because of the network proximity, cloudlet can offer a wireless connection with low latency and high bandwidth between the server and the CV, making it an ideal place for providing location-awareness, latency-intensive, and fast mobility management services and applications.

\emph{Fog computing} is another potential solution to address the presented issues in MCC. It is first proposed by Cisco in $2012$ \cite{bonomi2012fog}. The definition of fog computing is ``a system-level horizontal architecture that distributes resources and services of computing, storage, control, and networking anywhere along the continuum from cloud to Thing" \cite{openfog2016openfog}. Fog computing offers several compelling features \cite{bonomi2014fog} that are described below. (i) Heterogeneity, fog computing may contain a wide variety of computing nodes such as access points (APs), high-end servers, edge routers, RSUs, and even mobile nodes (e.g., smartphones and CVs). (ii) Geo-distribution and decentralized management, in contrast to cloud computing, fog computing is deployed widely geographical distributed at the edge of networks and manages its computation and storage resources in a decentralized way. (iii) Support for interplaying with the cloud, a cloud server is deployed at the top of the fog layer for deep analytics, where not only delay-intensive applications can be supported at the fog layer, but also computation-intensive and delay-tolerant applications (e.g., Big Data) can be performed at the cloud layer because of its large storage and powerful computing capability. Therefore, unlike MEC, fog computing often serves as a complement to a cloud rather than a substitute (i.e., fog computing ``cannot operate in a standalone mode'' \cite{mouradian2017comprehensive}). In Table \ref{Table2}, we present a comparison among different computing paradigms that we introduced above in terms of multiple key features, including architecture, location of computation resources, operation mode, etc.

Existing studies on CVs have proposed several cloud/edge/fog computing architectures based on the special requirements of different services/applications (will be discussed in Section \ref{sec:Lit_Survey}). Different cloud/edge/fog computing architectures may be suitable for supporting different types of CV applications with highly different Quality-of-Service (QoS) requirements in terms of latency, computation resources, and storage capacity. Under different architectures, the computing and communication workload for CVs may also vary over time and locations, which poses challenges to capacity planning, resource management of computation nodes, and mobility management of CVs. Thus, a well-designed computing architecture is very important for CV systems.

\subsection{Existing Surveys and Tutorials}

There are several related survey articles that focus on various aspects of cloud/edge/fog computing and CVs. We divide these existing survey papers into three categories: work on (i) MCC/edge/fog computing, (ii) vehicular networks, and (iii) MCC/edge/fog computing for CVs. In Table \ref{Table3}, we summarize published surveys on MCC/edge/fog computing. These articles focus on a wide range of issues related to MCC/edge/fog computing, including applications, architectures, computation offloading, taxonomy, security, standardization, communication, caching, resource management, and energy efficiency. However, none of them investigate the MCC/edge/fog computing for CVs. In Table \ref{Table4}, we summarize published surveys on vehicular networks, e.g., vehicular ad-hoc networks (VANETs), which discuss CVs only from the perspective of the communication.

Articles listed in Table \ref{Table5} are most related to our work, which discuss several research issues in the MCC/edge/fog computing for CVs. However, (i) the number of published surveys is quite few; (ii) these studies need to investigate more the system architecture design, where they only focus on one specific computing architecture in their whole paper, such as vehicular cloud computing (VCC) or vehicular fog computing (VFC); and (iii) to the best of our knowledge, there is no survey work that compares different architecture alternatives based on cloud/edge/fog computing for CVs or discusses their benefits, research challenges, and system requirements. Therefore, in view of prior survey work, there still lacks a systematic survey article offering comprehensive and concrete discussions on the architectural design alternatives based on cloud/edge/fog computing for CVs, which motivates this work.

\begin{table*}[tp!]
\newcommand{\tabincell}[2]{\begin{tabular}{@{}#1@{}}#2\end{tabular}}
\centering
\caption{Summary of Existing Survey Papers on MCC/Edge/Fog Computing}
\begin{tabular}{|c|c|c|l|}
\hline
{\textbf{Category}} & \textbf{Aspects} & {\textbf{Reference}} & {\textbf{Main Contribution}}\\\hline

\multirow{50}{*}{\tabincell{c}{MCC/Edge/Fog\\ Computing}} & \multirow{10}{*}{MCC} & \cite{othman2013survey} & \tabincell{l}{A comprehensive survey of MCC application models.}\\\cline{3-4}

&  & \cite{guan2011survey} & \tabincell{l}{A summary of challenges of MCC service designing, and a survey of MCC architecture, application\\ partition and offloading, and context-aware services.}\\\cline{3-4}

&  & \cite{dinh2013survey,patidar2012survey} & \tabincell{l}{An overview of the definition, applications, and architectures of MCC, along with the generic issues\\ and existing solutions. A discussion of the future research directions of MCC.}\\\cline{3-4} 

&  & \cite{fan2011survey} & \tabincell{l}{A discussion of exiting work on representative platforms and intelligent MCC access schemes.}\\\cline{3-4} 

&  & \cite{fernando2013mobile} & \tabincell{l}{A detailed taxonomy of MCC based on the key issues and the promising solutions to address them.}\\\cline{3-4} 

&  & \cite{alizadeh2016authentication} & \tabincell{l}{A comprehensive survey of the state-of-the-art authentication mechanisms in MCC.}\\\cline{3-4} 

&  & \cite{ramachandra2017comprehensive} & \tabincell{l}{A survey of security issues in MCC.}\\\cline{3-4} 

&  & \cite{varghese2018next} & \tabincell{l}{A discussion of next generation cloud computing in terms of research directions.}\\\cline{3-4} 
\cline{2-4}

& \multirow{22}{*}{\tabincell{c}{Edge\\Computing}} & \cite{mao2017survey} & \tabincell{l}{A comprehensive survey of the state-of-the-art MEC research from the communication perspective\\ with a focus on joint radio-and-computational resource management.}\\\cline{3-4}

&  & \cite{mach2017mobile} & \tabincell{l}{A comprehensive survey of major use cases and
reference scenarios, current advancement in\\ standardization of MEC, and research on computation offloading.}\\\cline{3-4}

&  & \cite{ai2018edge,abbas2017mobile,ahmed2017mobile,bilal2018potentials,taleb2017multi,shi2016edge} & \tabincell{l}{A comprehensive tutorial of three state-of-the-art edge computing technologies: MEC, cloudlet,\\ and fog computing. A comparison of standardization efforts, architectures, applications, and\\ principles, for these three technologies, as well as differences between MEC and\\ fog computing from the perspective of RANs.}\\\cline{3-4}

&  & \cite{beck2014mobile} & \tabincell{l}{A classification of applications deployed at the mobile edge according to the technical metrics\\ and the benefits of MEC for stakeholders in the network.}\\\cline{3-4}

&  & \cite{roman2018mobile} & \tabincell{l}{A discussion of the security threats and challenges in the edge paradigms, as well as the promising\\ solution for each specific challenge.}\\\cline{3-4}

&  & \cite{liu2019survey} & \tabincell{l}{A comprehensive overview of the existing edge computing systems and representative projects.\\ A comparison of open-source tools according to their applicability. }\\\cline{3-4}

&  & \cite{wang2017survey} & \tabincell{l}{A comprehensive survey of the state-of-the-art mobile edge networks with a focus on issues in\\ computing, caching, and communication techniques.}\\\cline{3-4}

&  & \cite{tocze2018taxonomy} & \tabincell{l}{A taxonomy for management and optimization of multiple resources in MEC.}\\\cline{3-4}

&  & \cite{khan2019edge} & \tabincell{l}{A classification of multi-facet computing paradigm within edge computing and identification of key\\ requirements to envision edge computing domain.}\\\cline{3-4}

&  & \cite{li2018edge} & \tabincell{l}{A comprehensive survey of edge-oriented computing systems with a focus on their architecture\\ features, management approaches, and design objectives.}\\\cline{3-4}

&  & \cite{olaniyan2018opportunistic} & \tabincell{l}{A presentation of the definition, computing paradigm, management framework, and research\\ challenges of the opportunistic edge computing.}\\\cline{3-4}

&  & \cite{porambage2018survey} & \tabincell{l}{A comprehensive survey of the realization of internet of things (IoT) applications within MEC.}\\\cline{3-4}

\cline{2-4}

& \multirow{15}{*}{\tabincell{c}{Fog\\Computing}} & \cite{mouradian2017comprehensive} & \tabincell{l}{A comprehensive survey on state-of-the-art fog computing from the perspective of architecture\\ and algorithm.}\\\cline{3-4}

&  & \cite{vaquero2014finding} & \tabincell{l}{An overview of the concept of fog computing in terms of enabling technologies and emerging\\ trends in usage patterns.}\\\cline{3-4}

 &  & \cite{yi2015survey} & \tabincell{l}{An overview of the definition of fog computing, representative application scenarios, and various\\ aspects of system issues.}\\\cline{3-4}
 
 &  & \cite{yannuzzi2014key,dastjerdi2016fog,bellavista2018survey,lin2017survey,atlam2018fog,sciarrone2016smart,elazhary2018internet,bittencourt2018internet} & \tabincell{l}{A discussion of the challenges of designing fog computing systems in IoT scenarios.} \\\cline{3-4}
 
 &  & \cite{zhang2018survey} & \tabincell{l}{An overview of access control of users' data in the environment of fog computing with the aim\\ of security challenges.}\\\cline{3-4}

 &  & \cite{yousefpour2019all,hu2017survey,luan2015fog} & \tabincell{l}{A comprehensive survey of fog computing, as well as its related computing paradigms, and a\\ detailed taxonomy of research topics in fog computing, including architecture, key technologies,\\ and applications.}\\\cline{3-4} 
 
 &  & \cite{elazhary2018internet,Dolui_2017_GIoTS} & \tabincell{l}{A comparison of fog computing, cloudlet and MEC and a discussion of their recommended use\\ cases.}\\\cline{3-4} 
 
 &  & \cite{yi2015security,zhang2018security} & \tabincell{l}{An overview of security and privacy issues of fog computing and a survey of existing solutions.}\\\cline{3-4} 
 
 &  & \cite{mukherjee2018survey} & \tabincell{l}{A comprehensive survey of fog computing from the network perspective and a discussion of\\ several network issues, such as latency, bandwidth, and energy consumption in fog computing.}\\\cline{3-4} 
\hline

\end{tabular}
\label{Table3}
\end{table*}

\begin{table*}[tp!]
\newcommand{\tabincell}[2]{\begin{tabular}{@{}#1@{}}#2\end{tabular}}
\centering
\caption{Summary of Existing Survey Papers on Vehicular Networks}
\begin{tabular}{|c|c|c|l|}
\hline
{\textbf{Category}} & \textbf{Aspects} & {\textbf{Reference}} & {\textbf{Main Contribution}}\\\hline

\multirow{16}{*}{Vehicular networks} & \multirow{1}{*}{Taxonomy} & \cite{cooper2016comparative}& \tabincell{l}{A taxonomy of the techniques applied to solve the issues of VANET cluster head election,\\ cluster affiliation, and cluster management, as well as a discussion of research directions\\ and trends in the design of these algorithms.}\\\cline{3-4}
\cline{2-4}

& \tabincell{c}{CV} & \cite{rios2016survey} & \tabincell{l}{A summary of the state-of-the-art developments and the research trends in coordination\\ with the connected and automated vehicles.}\\\cline{3-4}
\cline{2-4}

& \multirow{1}{*}{\tabincell{c}{Routing}} & \cite{li2007routing,altayeb2013survey,bilal2013position,awang2017routing} & \tabincell{l}{A classification of existing routing protocols developed for VANETs, as well as\\ comparisons among different classes.}\\\cline{3-4}
\cline{2-4}

& \tabincell{c}{Mobility\\management} & \cite{zhu2011mobility} & \tabincell{l}{A comprehensive survey of mobility management for vehicular networks, including the\\ design requirements and existing solutions based on V2V and V2I.}\\\cline{3-4}
\cline{2-4}

& \multirow{5}{*}{\tabincell{c}{Wireless\\technologies\\ \& protocols}} & \cite{al2014comprehensive,zheng2015heterogeneous} & \tabincell{l}{A survey of wireless access technologies, characteristics, challenges, and requirements in\\ VANET, as well as a summary of simulation tools and models of VANET.}\\\cline{3-4}
&   & \cite{lu2014connected,zheng2015heterogeneous} & \tabincell{l}{An overview of the state-of-the-art wireless solutions for vehicle-to-sensor, V2V,\\ vehicle-to-Internet, and vehicle-to-road infrastructure connectivities.}\\\cline{3-4}
&   & \cite{cunha2016data,zheng2015heterogeneous,hartenstein2008tutorial,karagiannis2011vehicular} & \tabincell{l}{A comprehensive survey of wireless communication protocols, standards, architectures, and\\ applications of VANETs.}\\\cline{3-4}

\cline{2-4}

& \multirow{1}{*}{\tabincell{c}{Security}} & \cite{othmane2015survey} & \tabincell{l}{A taxonomy of security and privacy aspects of CV, including security of communication\\ links, data validity, security of devices, identity and liability, access control, and privacy\\ of drivers and vehicles, as well as existing solutions.}\\\cline{3-4}
\cline{2-4}

& \tabincell{c}{Green\\ networks} & \cite{alsabaan2012vehicular} & \tabincell{l}{A discussion of green vehicular networks design, including the communication protocol,\\ routing protocol, mobility models, and open issues.}\\\cline{3-4}

\hline

\end{tabular}
\label{Table4}
\end{table*}

\begin{table*}[tp!]
\newcommand{\tabincell}[2]{\begin{tabular}{@{}#1@{}}#2\end{tabular}}
\centering
\caption{Summary of Existing Survey Papers on MCC/Edge/Fog Computing for CV}
\begin{tabular}{|c|c|c|l|}
\hline
{\textbf{Category}} & \textbf{Aspects} & {\textbf{Reference}} & {\textbf{Main Contribution}}\\\hline

\multirow{10}{*}{\tabincell{c}{MCC/Edge/Fog\\ computing for CV}} & {\tabincell{c}{Vehicular cloud\\ computing}} & \cite{whaiduzzaman_2014,boukerche2018vehicular,Gu_2013_GC_Wkshps}& \tabincell{l}{A comprehensive survey of VCC, including inter-cloud communication systems, featuring\\ applications, services, cloud formations, traffic models, key management, and security issues.}\\\cline{3-4}

\cline{2-4}& \multirow{3}{*}{\tabincell{c}{Vehicular fog\\ computing}} & \cite{menon2017moving} & \tabincell{l}{A discussion of challenges and future trends of vehicular fog computing.}\\\cline{3-4}
&  & \cite{huang2017vehicular} & \tabincell{l}{A presentation of a high-level system architecture and a typical use case in vehicular fog\\
computing, as well as security and forensic challenges and potential solutions.}\\\cline{2-4}

\cline{2-4}& \multirow{6}{*}{\tabincell{c}{Vehicular\\ applications}} & \cite{giang2016developing} & \tabincell{l}{An investigation on how smart transportation applications are developed following fog\\ computing along with their challenges and possible mitigation from the state of the arts.}\\\cline{3-4}
&  & \cite{grover2018real} & \tabincell{l}{A discussion on how real-time VANET applications can be developed in fog computing\\ systems.}\\\cline{3-4}
&  & \cite{Bitam_2015_IWC} & \tabincell{l}{A proposal for a new cloud computing model, VANET-cloud, to improve traffic safety and\\ provide computation services for road users, as well as an overview of some future research\\ directions, including security, energy efficiency, data aggregation, resource management, and\\ interoperability.}\\\cline{3-4}
\hline

\end{tabular}
\label{Table5}
\end{table*}

\subsection{Contribution}

In contrast to the above-mentioned surveys, this paper provides a comprehensive survey on different architectural design alternatives based on cloud/edge/fog computing for CVs. The main contributions of this paper are presented as follows:

\begin{itemize}
\item Presenting a holistic overview on the design of CV systems from both academia and industry perspectives, including activities in industry, functional requirements (Section \ref{sec:Con_Veh}), service requirements, and design considerations (Section \ref{sec:Requirements}).

\item Providing a comprehensive literature survey on existing proposed architectural design alternatives based on cloud/edge/fog computing for CVs (Section \ref{sec:Lit_Survey}).

\item Proposing a new classification of computing architectures based on cloud/edge/fog computing for CVs: computation-aided and computation-enabled architectures (Section \ref{sec:Lit_Survey}).

\item Presenting a holistic comparison among different cloud/edge/fog computing architectures for CVs based on functional requirements of CV systems, including advantages, disadvantages, and research challenges (Section \ref{sec:Lit_Survey}).

\item Proposing several open research issues of designing cloud/edge/fog computing architectures for CVs, including other hybrid architectural alternatives, localizing data traffic, mobility support in heterogeneous architectures, and computing resource management (Section \ref{sec:openissue}).
\end{itemize}

\subsection{Paper Organization}
The rest of the paper is organized as follows: In Section \ref{sec:Con_Veh}, we first present an overview introduction on the design of CV systems with a brief summary on the activities of the U.S. Department of Transportation on CVs. The main functions of a CV eco-system are also described. In Section \ref{sec:Requirements}, we summarize the service requirements and design considerations of using cloud/edge/fog computing for CV applications. Existing architectural design alternatives in the literature, i.e., computation-aided computing and computation-enabled computing, are holistically surveyed and compared in Sections \ref{sec:Lit_Survey}. Open research issues are discussed in Section \ref{sec:openissue}. Finally, we conclude in Section \ref{sc:conclusion}. Table \ref{Table6} presents the list of acronyms used in this paper.

\begin{table}[tp!]
\newcommand{\tabincell}[2]{\begin{tabular}{@{}#1@{}}#2\end{tabular}}
\centering
\caption{List of acronyms and their descriptions}
\begin{tabular}{|l|l|l|l|l|l|l|}
\hline
Acronym        & Description  \\\hline
 5G-PPP & 5G Public Private Partnership \\\hline
 AECC & Automotive Edge Computing Consortium \\\hline
 AP  &  Access Point \\\hline
 ARC-IT & \tabincell{l}{Architecture Reference for Cooperative and Intelligent \\Transportation} \\\hline
 BS  &  Base Station \\\hline
 BASN & Body  Area  Sensor  Network  \\\hline
 CV  &  Connected Vehicle  \\\hline
 CTS & Clear-to-Send \\\hline
 CSMA/CA & Carrier Sense Multiple Access with Collision Avoidance \\\hline
 CPU & Central Processing Unit \\\hline
 C-ITS & Cooperative Intelligent Transport System \\\hline
 DoS & Denial of Service \\\hline
 D2D & Device-to-Device \\\hline
 DSRC & Dedicated Short-range Communication \\\hline
 ETSI & European Telecommunications Standards Institute \\\hline
 FCC & Federal Communications Commission \\\hline
 GPS & Global Positioning System \\\hline
 HOV & High Occupancy Vehicle \\\hline
 IT  &  Information Technology      \\\hline
 IoT & Internet of Things \\\hline
 IaaS & Infrastructure-as-a-Service \\\hline
 \multirow{1}{*}{ITS-G5} & \tabincell{l}{Intelligent Transport Systems operating in the 5 GHz\\ frequency band} \\\hline
 LOS & Line-of-Sight \\\hline
 LTE & Long-Term Evolution \\\hline
 MAN & Metropolitan Area Network \\\hline
 MIMO & Multiple-Input Multiple-Output\\\hline
 MU & Mobile User \\\hline 
 MCV & Maintenance and Construction Vehicle \\\hline
 MCC &  Mobile Cloud Computing     \\\hline
 MEC &  Mobile Edge Computing/Multi-Access Edge Computing \\\hline
 MAC & Medium Access Control \\\hline
 OTA & Over-the-Air \\\hline
 OEM &  Original Equipment Manufacturer\\\hline
 PKI & Public Key Infrastructure  \\\hline
 PaaS & Platform-as-a-Service \\\hline
 QoS &  Quality of Service      \\\hline
 RAN & Radio Access Network \\\hline
 RSU & Roadside Unit \\\hline
 RSE & Roadside Equipment \\\hline
 RTS & Request-to-Send \\\hline
 RTT & Round-Trip Time \\\hline
 SCMS & Security Credential Management System \\\hline
 SDN & Software Defined Network \\\hline
 SaaS & Software-as-a-Service \\\hline
 TMC & Transportation Management Center \\\hline
 TDMA & Time Division Multiple Access \\\hline
 USDOT & United States Department of Transportation \\\hline
 V2X &  Vehicle-to-everything    \\\hline
 V2V &  Vehicle-to-Vehicle       \\\hline
 V2I &  Vehicle-to-Infrastructure \\\hline
 V2C &  Vehicle-to-Cloud    \\\hline
 V2D &  Vehicle-to-Driver \\\hline
 VANET & Vehicular ad-hoc Network \\\hline
 VC & Vehicular Cloud \\\hline
 VCC & Vehicular Cloud Computing \\\hline
 VC-MAC & Vehicular Cooperative Media Access Control \\\hline
 VFC & Vehicular Fog Computing \\\hline
 VM & Virtual Machine \\\hline
 WLAN & Wireless Local Area Network  \\\hline
 WiMAX & Worldwide Interoperability for Microwave Access \\\hline
 WiBro & Wireless Broadband \\\hline
 WAN & Wide Area Network \\\hline
 WAVE & Wireless Access in Vehicular Environments \\\hline
\end{tabular}
\label{Table6}
\end{table}

\section{CV System Design}\label{sec:Con_Veh}
Before diving into the computing architectures for CVs, we first give an overview introduction on the design of CV systems.
In particular, we first introduce some CV projects initiated by the U.S. government and European Commission. Then, we summarize the functional requirements of a CV eco-system.

The United States and Europe advances on the deployment of CVs are summarized in paper \cite{uhlemann2017us}. In particular, the United States Department of Transportation (USDOT) issued a new rule in December 2016 that requires that V2V technologies must be implemented in all the new manufactured light-duty vehicles. Thus, developing standardized messaging technology together with industry can efficiently improve the deployment of CV technologies in the U.S. In addition, the U.S. version of IEEE 802.11p and the dedicated short-range communications (DSRC) are the two alternatives for transmitting data (e.g., vehicle speed, direction, and location) among vehicles using V2V communications. Therefore, V2V-equipped vehicles can identify risks and provide warnings to drivers to avoid imminent crashes. Other activities initiated by USDOT are explained in the following sub-section.

Similarly, the European Commission submitted the European Strategy on Cooperative Intelligent Transport Systems (C-ITS) in November 2016. C-ITS messages will be transmitted for a wide range of services between different vehicles. To support all C-ITS services on the vehicle side, a full hybrid communication mix needs to be on board. Currently, the commission considers a combination of ETSI ITS-G5 (The European Telecommunications Standards Institute, Intelligent Transport Systems operating in the 5 GHz frequency band), the European version of IEEE 802.11p, and existing cellular networks as the promising hybrid communication mix that ensures the best possible support for deploying C-ITS services.

\subsection{USDOT Activities on CVs}
\label{sec:USDOT}

The USDOT initiated many CV projects by interacting with a wide range of stakeholders. One of the projects is the CV pilot projects~\cite{USDOT-Pilot} launched in three different regions in U.S.A. --- Wyoming, New York city, and Tampa. The main purpose of this project is to demonstrate how to improve driving safety and comfort by allowing vehicles to communicate with road-side units or centers; those applications may include, but not limited to, pedestrian collision avoidance, early warning of severe weather conditions, traffic flow improvement, and so on\footnote{USDOT listed more than one hundred potential connected applications in Architecture Reference for Cooperative and Intelligent Transportation (ARC-IT) website~\cite{ARC-IT} and some of them are the target applications of these pilot projects.}.

\begin{figure}[tp!]
\centering
\includegraphics[width=0.48\textwidth]{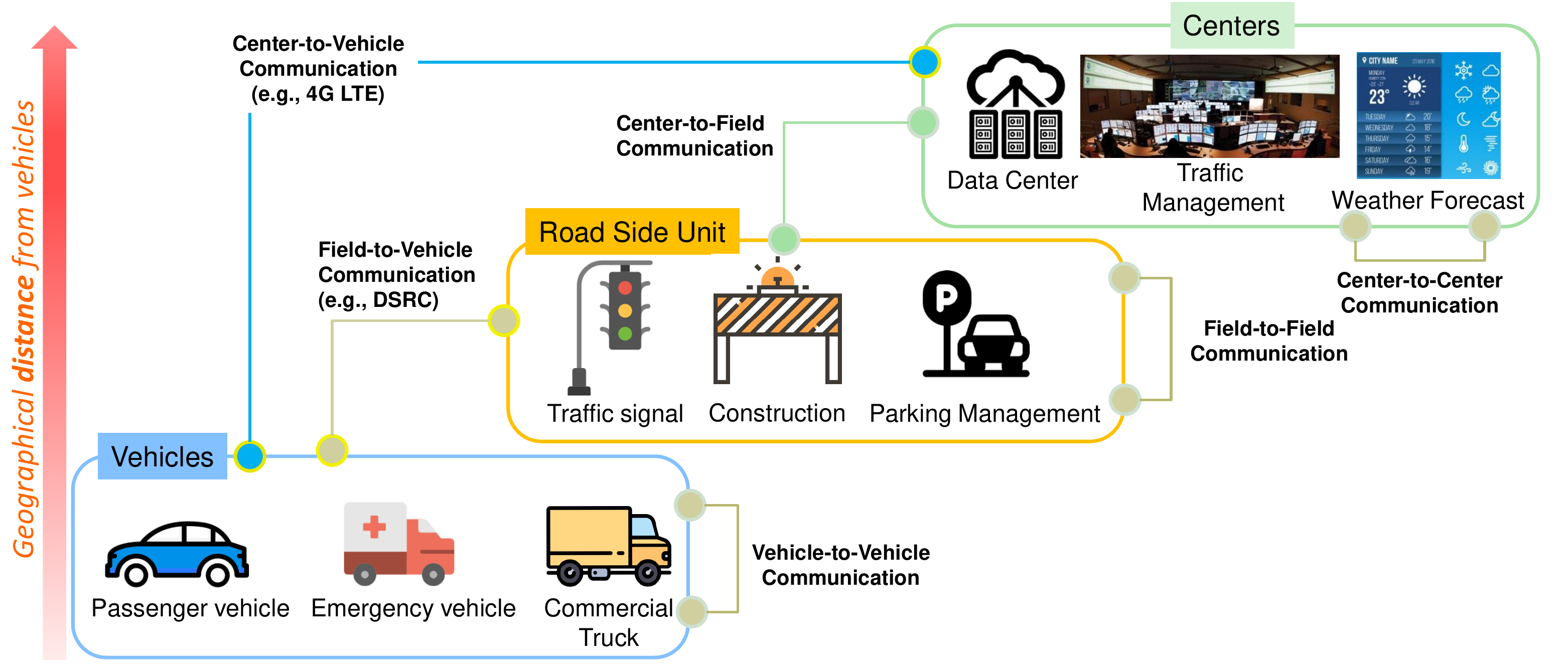}
\caption{The architecture overview of the CV eco-system.}
\label{Figure1}
\end{figure}

USDOT is developing various open reference system architectures~\cite{ARC-IT} that are specific to implement specific use cases, but the common aspects of those architectures can be summarized as shown in Fig. \ref{Figure1}. This high-level architecture illustrates the overview of the CV eco-system that consists of three levels: vehicles, RSUs, and centers. The corresponding communications in such an eco-system include intra-level communications (i.e., V2V communication, field-to-field communication (from one RSU to another), and center-to-center communication) and inter-level communications (i.e., field-to-vehicle communication, center-to-vehicle communication, and center-to-field communication).

Various types of vehicles are considered in this eco-system, such as passengers’ vehicles, emergency vehicles, or trucks that have wireless communication capability. Those vehicles may communicate with the RSUs installed in the close proximity of the vehicles. RSUs are typically equipped with computation units that can perform local computation and wireless communication that allow them to exchange messages with vehicles or other systems. The general role of RSUs is to make a local decision based on the data collected from vehicles or centers, but their specific roles may vary depending on the applications. In New York city pilot project~\cite{NY-Pilot}, for example, the RSUs are installed in urban intersection areas so that it can monitor pedestrian crossing or approaching vehicles and send warning messages to them; in Wyoming city pilot project~\cite{WY-Pilot}, the RSUs monitor the hazardous weather or road conditions on the rural highway via on-board sensors and inform any necessary warnings via wireless communications.

Centers are the largest system that can monitor the data gathered from vehicles or RSUs and perform global decisions that can make a broader impact on the overall eco-system. Such a decision may include to control a range of deployed RSUs or to provide generic traffic services to vehicles.

Regarding security, USDOT designs the Security Credential Management System (SCMS)~\cite{USDOT-SCMS} which is a proof-of-concept security solution for CVs. The SCMS is based on Public Key Infrastructure (PKI), and its goal is to ensure integrity, authenticity, and privacy for the communication between CVs, RSUs, and aftermarket safety devices.

\subsection{Functional Requirements of CV Systems}\label{sec:Functions}
Now, we briefly explain the five main functions of such a CV eco-system described above.

\textit{Data Sharing}: Data sharing is when vehicles share their collected data in CV systems. The process of data sharing can be generally divided into three levels: V2V (e.g., in a VANET), V2D (e.g., in a body area sensor network (BASN)), and V2I. Different services may require varying size of collected data. The bandwidth required for data sharing and the throughput of data sharing are different in the three different data sharing levels.

\textit{Data Processing}: Data processing is when computing units (e.g., the centers, RSUs, and vehicles) process the collected data. For example, in the intelligent driving scenario, collected data from vehicles such as cruising, video, and control data need to be offloaded to a computing unit which then processes these data heavily. Thus, the computing power of computing units constrains what type of services they can support.

\textit{Monitoring}: Monitoring is when upper-level entities in the system monitor the presence and experience of lower-level entities. For example, the data center monitors the presence and experience of RSUs, or a field element monitors the vehicle presence and experience.

\textit{Warning}: Warning is when the center, field equipment, or vehicles offer advisories and warnings to drivers, such as current road conditions and predicted weather events.

\textit{Control}: Control is when upper-level entities in the system send control instructions to lower-level entities. For example, an RSU checks a vehicle's condition to see if it is suitable for operating on automated lanes based on certain vehicle control parameters (such as speed and headway) that will be used by the vehicle in those lanes, and once confirmed, sends the control parameters to the vehicle.

Note that every function is possibly associated with its required security levels. In USDOT ARC-IT, required security levels of confidentiality, integrity, and availability are provided for physical objects and information flows.

In Table \ref{Table7}, under different example applications of CVs, the requirements of these five functions at different levels of the eco-system as illustrated in Fig. \ref{Figure1} are listed.

\begin{table*}[tp!]
\newcommand{\tabincell}[2]{\begin{tabular}{@{}#1@{}}#2\end{tabular}}
\centering
\caption{Example Functional Requirements of CVs \cite{USDOT-SP}}
\begin{tabular}{|c|c|l|}
\hline
{\textbf{Category}} & \textbf{Sub category} & \textbf{Example requirements} \\\hline
\multirow{13}{*}{Data Sharing} & Center-to-Center & \tabincell{l}{CVO10-10 (Fleet Administration): Centers shall share data on current road conditions and predicted weather\\ events with each other.} \\
\cline{2-3}& Vehicle-to-Center & \tabincell{l}{CVO01-02 (CV On-Board Trip Monitoring): A CV shall offer the route details to the driver as received\\ from the fleet management center.} \\
\cline{2-3}& Field-to-Center & \tabincell{l}{CVO10-05 (TMC Environmental Monitoring): A field element, such as the CV roadside equipment, shall\\ send the aggregated and processed vehicle environmental data that is collected through the vehicle safety\\ and convenience system to the center.} \\

\cline{2-3}& Field-to-Field & \tabincell{l}{CVO06-4 (RSE Intersection Management): A field element shall transmit the request for right-of-way such\\ as signal preemption and priority to a traffic signal controller.} \\
\cline{2-3}& V2V & \tabincell{l}{PS07-05 (Vehicle Basic Safety Communication): A CV shall exchange location and motion information with\\ roadside equipment and nearby CVs.} \\
\cline{2-3}& Vehicle-to-Field & \tabincell{l}{CVO06-02 (CV On-Board Signal Priority): A CV shall provide its location and motion information to local\\ CV equipment near signalized intersections.} \\
\cline{2-3}
\hline
\multirow{8}{*}{Data Processing} & Center-level & SU04-10 (Map Management): A center shall use CV location information to refine roadway geometry.\\
\cline{2-3}& Field-level & \tabincell{l}{PS03-07 (RSE Intersection Management): A field element shall decide when a special CV that requests signal\\ preemption or signal priority is approved based on its digital credential.} \\
\cline{2-3}& Vehicle-level & \tabincell{l}{CVO01-01 (CV On-Board Trip Monitoring): A CV shall compute the location of the vehicle itself and\\ its freight equipment based on multiple measured information (e.g.,identity and distance traveled) and\\ a positioning system.} \\
\cline{2-3}& Personal-device & \tabincell{l}{SU04-03 (Personal Map Management): A personal device shall make basemap, roadway geometry,\\intersection geometry, and parking facility geometry information available to other personal device\\ applications.} \\
\cline{2-3}
\hline
\multirow{5}{*}{Monitoring} & Center-level & \tabincell{l}{CVO01-18 (Fleet Administration): A center shall monitor the status of its fleet, including CVs and freight\\ equipment, for maintenance problems.} \\
\cline{2-3}& Vehicle-level & \tabincell{l}{MC07-01 (MCV Vehicle Safety Monitoring): A maintenance and construction vehicle shall monitor that if\\ a vehicle has entered a work zone without permission.} \\
\cline{2-3}& Field-level & \tabincell{l}{ST04-01 (RSE Lighting System Support): A field element shall monitor vehicle presence and collect \\environmental information from passing CVs.} \\
\cline{2-3}
\hline

\multirow{5}{*}{Warning} & Center-level & \tabincell{l}{CVO10-16 (Fleet Administration): A center shall broadcast current road condition and predicted weather\\ event warnings to drivers.} \\
\cline{2-3}& Field-level & \tabincell{l}{PS07-01 (RSE Incident Scene Safety): A field equipment shall provide CVs notification of an \\incident scene emergency or safety issue.} \\
\cline{2-3}& Vehicle-level & \tabincell{l}{CVO01-03 (CV On-Board Trip Monitoring): A CV shall warn the commercial vehicle fleet management\\ center when the vehicle's location has deviated from its planned route.} \\
\cline{2-3}
\hline
\multirow{3}{*}{Control} & Center-level & \tabincell{l}{CVO06-10 (TMC Signal Control): A center shall adjust signal timing according to a signal preemption, \\signal prioritization, multi-modal crossing activation, pedestrian call, or other requests for right-of-way.} \\
\cline{2-3}& Field-level & \tabincell{l}{MC03-02 (Roadway Automated Treatment): A field element shall activate automated roadway \\treatment systems (e.g., anti-icing, fog dispersion, and chemicals) under the center control.} \\
\cline{2-3}
\hline

\end{tabular}
\label{Table7}
\end{table*}

\section{Service Requirements and Design Considerations of Cloud/Edge/Fog Computing for CVs}\label{sec:Requirements}

To realize such a CV eco-system with the five main functions as explained in Section \ref{sec:Con_Veh}, different computing architectures can be considered. One alternative is to adopt a cloud-based computing architecture where the centers in Fig. \ref{Figure1} are located in a remote cloud. Another alternative is to consider an edge/fog-based computing architecture to bring the computing capabilities closer to vehicles or field equipment. In this way, the centers in Fig. \ref{Figure1} are distributed at multiple locations in the system. In this section, we first introduce the service requirements in Section \ref{ssc:Service Requirements}, and then the design considerations to realize a cloud/edge/fog computing system for CVs in Section \ref{ssc:Design Considerations}.

\subsection{Service Requirements}
\label{ssc:Service Requirements}

There are a wide range of IoT devices that provide many different types of services. Some may share commonality with CV services, while some are not. The purpose of this subsection is to highlight the unique characteristics of CV services via comparison with a representative IoT device. For this comparison, we chose a smartphone as a comparison peer, because (1) its user base is as wide as the one of vehicles, (2) continuous connectivity is required for most applications, (3) and it has a mobility aspect that a user is expected to receive services while moving. Even though the uniqueness claimed in this section may not be generalized across all other IoT devices beyond smartphones, we believe this gives a good insight as to the major challenges in realizing CV services.

\subsubsection{Data Generation in vehicular networks}
In comparison to smartphones, the amount of data generated from a vehicle is huge in its volume. A high-end vehicle is typically equipped around a hundred sensors or more to monitor correct system operation, and enhance safety and driving comfort. Even though not all raw sensor data need to be transferred to the remote cloud, it is generally expected that each vehicle needs to send at least 20 GB of data per month to the cloud to achieve practical automotive applications according to Automotive Edge Computing Consortium (AECC) \cite{AECC_2018_Whitepaper}; in contrast, in spite of varying statistics, a typical smartphone user consumes around 2 to 5 GB of cellular data these days\footnote{These statistics exclude Wi-Fi usages.}.

In addition, the dataflow pattern of vehicle data is quite different from smartphones. Most smartphones are dedicated for downloading services; that is, the remote cloud server is typically a data producer that creates contents, and sends it to the smartphone, which is a data consumer. Due to this characteristic, many Internet Service Providers assign higher network bandwidth to the downlink services than uplink services. On the other hand, a vehicle is more likely to become a data producer, which generates various raw data, and send it to the cloud for being used by additional services.

\subsubsection{Response Time}
One typical type of the service response time is a duration from the moment a user requests a data or computation to the moment it has been completed or received by the user. Both smartphone and CV services need to meet diverse granularity of timing requirements; for example, real-time multi-user game (smartphone service) or road-side object recognition (CV service) typically need to meet the response time in the order of milliseconds; on the other hand, storage backup application (smartphone service) or HD map generation (CV service) need to meet the response time in the order of seconds or minutes.

However, the consequences of violating such expected response time is significantly different each other, so designing those systems also become different. Many CV services are safety-critical services where delayed response time has a safety impact on drivers or others. For example, a vehicle platooning service that needs to guarantee a constant distance among a group of vehicles may end up crashing each other unless a series of positions of other vehicles do not arrive on time. For this reason, many CV services are typically equipped with a fail-safe mode that is activated when such abnormal condition arises. Hence, the architecture should be designed more robustly to cope with such abnormal delays and to provide sufficient information to activate such a fail-safe mode. On the other hand, most smartphone applications do not have safety implication on the users when the response time is delayed, so the supporting architecture typically do not accompany with such a consideration in place.

\subsubsection{Availability-Cost Tradeoff}

Both smartphone and vehicles may be equipped with various services that require different degrees of network availability depending on their service requirements. Some services require high network availability to provide proper functionalities such as video streaming (smartphone service) or vehicle platooning (CV service). On the other hand, other services may only require intermittent network availability as their local compute unit and storage can support the continuation of the services without continuous network connectivity, such as downloadable standalone games (smartphone service), downloadable navigation map (CV service).

Even though there is an ongoing debate as to the best way to provide connectivity for future vehicles \cite{mahmood2019software,xu2017dsrc,abboud2016interworking}, vehicles may be exposed more heterogeneous wireless networks that have different costs and latencies than smartphones while they are moving; a cellular network is typically the only option for a smartphone to maintain the connectivity while it is moving at a similar speed with a vehicle. In United States, some of ongoing V2I services \cite{USDOT-Pilot} are provided by government via DSRC that allows DSRC-compatible vehicle to freely receive the public services. At the same time, a vehicle is also equipped with a cellular modem that can transfer other types of data via cellular network, which incur costs in most cases. This requires a moving vehicle to make a unique design decision, which does not arise in a moving smartphone, as to when multiple network options are available on the vehicle route, how to schedule the service execution by considering various aspects, such as latency, cost and so on.

\subsubsection{Data Security and Privacy}
Unlike smartphones’ impact resulted from security or privacy attack, the CV service attack has safety implications as those services are linked to safety-critical control applications. For example, a roadway signal infrastructure may broadcast safety messages (e.g., pedestrian positions or speed limits) in intersections for a crash-mitigation service \cite{USDOT-Pilot}; when the information is compromised by attackers, a vehicle that utilizes the fake information may trigger unexpected control operation resulting in safety issues such as unexpected hard braking due to fake pedestrian crossing information or unexpected speed increase due to fake speed limit. 

However, it is also challenging to achieve the necessary degree of security and privacy as it typically negatively affects system performance and convenience. For example, adding strong encryption to all data from cloud or out from vehicles may add extra complexity to the vehicle system design such as latency, extra compute and storage power. Therefore, it is necessary to impose different types of security measure depending on their criticality levels accounting for their interaction with control-related systems. Note that a smartphone also provides multi-level security and privacy measures, but the burden to achieve the required level of security and privacy is mainly imposed to the users by asking more credential (e.g., multi-factor authentication). However, it is not possible for drivers to follow such complex procedure during driving, so it is desirable to perform such procedure more seamlessly.

\subsubsection{Data Locality and Data Sovereignty}
Vehicle data has a higher locality than the one used for smartphone services. Many CV services utilize data that is only consumed in the areas where it is originating, such as positions of other vehicles, semantics of road signs, local HD map information and so on; that is, such data is meaningless in other remote areas irrelevant to the CV services. Therefore, it is not desirable to send all data to remote clouds as it consumes the network and compute resources unnecessarily such as network bandwidth or cloud storage. The system architecture should be able to support the unique characteristics of data locality for CV services so that the infrastructure-wide resources can be efficiently utilized for other non-CV services as well.

In addition, as data is increasingly an important asset to each country, it is necessary to follow the local rules and regulations imposed by each nation. For example, some countries may restrict some type of data to physically stay in their territories depending on how they are used. If a vehicle is used for a certain that falls in such a restrictive category, the data transmission should strictly follow the local regulation. These days, OEMs typically do not have that level of customization as it increases the manufacturing cost significantly. However, this situation will arise as more data is shared with remote cloud or vehicles, so it is necessary to consider a system architecture design to enable data to be transmitted conforming such local regulation via support from either in-vehicle system or infrastructure.

\subsection{Design Considerations}
\label{ssc:Design Considerations}
Given the service requirements described above, the challenges and considerations of cloud/edge/fog computing architectural design for CVs are discussed in this sub-section.

\subsubsection{Networking}
Due to the reason that vehicular connections are usually uncertain and frequently changing in topology, and thus the reliability of vehicular networks is still challenging. At the same time, the bandwidth resources of cellular networks are limited and BSs' signal cannot extend to all the urban and suburban areas. Therefore, we need to design a heterogeneous vehicular network that combines the best of cellular networks and V2V ad hoc networking. In such a heterogeneous vehicular network, resource sharing and co-scheduling among different networks is still an open issue. A lot of work, currently, has investigated in resource sharing in 5G-enabled vehicular networks \cite{Yu_2016_ITVT}. However, so far co-scheduling mechanism design is still lacking for CVs with heterogeneous communication network support\cite{Xiao_2017_IIC}.

\subsubsection{Data Sharing in Vehicular Networks}
Since co-located vehicles often require shared content, such as navigation or environment recognition, the broadcast nature of V2V communications can improve the performance of content sharing in vehicular networks. However, due to the fact that IEEE 802.11p utilizes the carrier sense multiple access with collision avoidance (CSMA/CA) mechanism, the request-to-send/clear-to-send (RTS/CTS) handshake is disabled in broadcast \cite{Q_2018_IN_Toward}. Thus, V2V communications are subject to a severe hidden terminal problem that will cause potential collisions at the receivers. Therefore, an efficient medium access control (MAC) mechanism should be designed to avoid such collisions. According to the space-constraint deadlines, the data demands have different urgent levels, even for the same content. Thus, in a certain area, sharing different data contents will have different gains. In order to increase the amount of data transmitted through vehicular networks as much as possible, three issues should be carefully planned: (i) which content to broadcast; (ii) when to broadcast; and (iii) which vehicle to broadcast. Edge servers that know the served vehicles' content demands and positions, can coordinate transmissions among vehicles to improve the network gain and make sure no collisions among the served vehicles. However, a single edge server cannot realize collision avoidance among vehicles in different locations. Timely and frequent interactions among multiple edge servers should be introduced to avoid such collisions. Furthermore, the frequent control messages among multiple edge servers and served vehicles will incur extra overhead for vehicular networks. Thus, it is necessary for vehicular networks to share contents in a distributive and cooperative manner \cite{Q_2018_IN_Toward}.

\subsubsection{Application Deployment}
Vehicular applications can be deployed in cloud centers, edge/fog nodes, or vehicles. The application deployment depends on factors such as the vehicular network topology, users' delay tolerance, and the vehicles' and users' mobility predictions. In paper \cite{Xuan_2016_APNOMS}, the authors focused on application deployment on the rented cloud nodes or the own fog nodes, and proposed a heuristic-based algorithm that tries to make a trade-off between the makespan and the expenditures of cloud nodes. In paper \cite{Huang_2016_Reliable}, taking the mobility of mobile devices into consideration, an adaptive content reservation scheme, which reserves the resources on the cloud centers and fog nodes for real-time video streaming to mobile devices, is proposed. In paper \cite{Lin_2015_ICPP}, the authors proposed to develop a new fog that treats the idle devices of game players or organizations as fog nodes, rendering game streaming to the nearby players. Still, it is challenging to consider the mobility of edge/fog nodes in application deployment. On the one hand, both the vehicles and the data sources may move at the same time. On the other hand, it is complex to coordinate the computation and communication systems simultaneously \cite{Xiao_2017_IIC}.

\subsubsection{Security and Privacy}
Most existing researches focus on the potential attacks or threats in cloud/edge/fog-assisted vehicular applications. Attacks can be generally classified into two types, active attacks and passive attacks. The functionality of a vehicular cloud/edge/fog system cannot be destroyed by passive attacks which only want to eavesdrop the private information. However, the active attacks are more damaging than the passive ones, due to the reason that the active ones attempt to interrupt the operations of the cloud/edge/fog computing systems, or modify the sharing data. Active attacks are usually easy to be detected when it induces huge damages to the vehicular system. However, it is hard to be noticed if the attacks are performed in an inconspicuous manner within a very limited time \cite{huang2017vehicular}. Additionally, the attackers generally falls into two categories: insider and external attackers. An insider attacker comes from inside the vehicular system, and are usually equipped with key materials. An internal attacker may induce more potential risks than an external attacker, since the internal attacker knows the existing security control policy and can circumvent it. 

Specifically, in paper \cite{Pra_Samuel_2015}, the authors show that in a connected car environment, a real vehicle and malicious smartphone application can be used to perform a long-range wireless attack. Then, a security protocol for controller area network is proposed as a countermeasure. In paper \cite{Security_Gong_2013}, the authors identify the security challenges that are specific to vehicular clouds (VCs), including authentication of high-mobility vehicles, scalability and single interface, the complexity of establishing trust relationships among multiple players caused by intermittent communications, and tangled identities and locations. A security scheme is proposed to address several of the aforementioned challenges. In paper \cite{Security_Haojin_2009}, security issues in service-oriented vehicular networks are elaborated, i.e., minimizing V2I authentication latency and distributed public key revocation. These two security issues are considered as among the most challenging design targets in service-oriented vehicular networks. Accordingly, a fast V2I authentication based vehicle mobility prediction scheme and an infrastructure-based short-time certificated management scheme are proposed to address the aforementioned two challenges. In paper \cite{huang2017vehicular}, the authors discuss several key security and forensic challenges and their potential solutions. A secure VFC implementation should provide multiple baseline security and forensic properties, including confidentiality, integrity, authentication, access control, non-repudiation, availability, reliability, and forensics. Most of the security requirements can be achieved by cryptographic techniques. Moreover, the authors investigate the compromise attack and selfish attacks, and their potential countermeasures. In paper \cite{roman2018mobile}, the authors holistically analyze the mechanisms, challenges, and security threats existing in all edge scenarios, while highlighting the collaboration and synergies among them. In paper \cite{Othmane2015}, the authors consider the security and privacy aspects of CVs, including security of communication links, data validity, security of devices, identity and liability, access control and privacy issues.

\section{Cloud/Edge/Fog Computing Architectures for CVs}

\label{sec:Lit_Survey}

\begin{figure*}[tp!]
\centering
\includegraphics[width=0.98\textwidth]{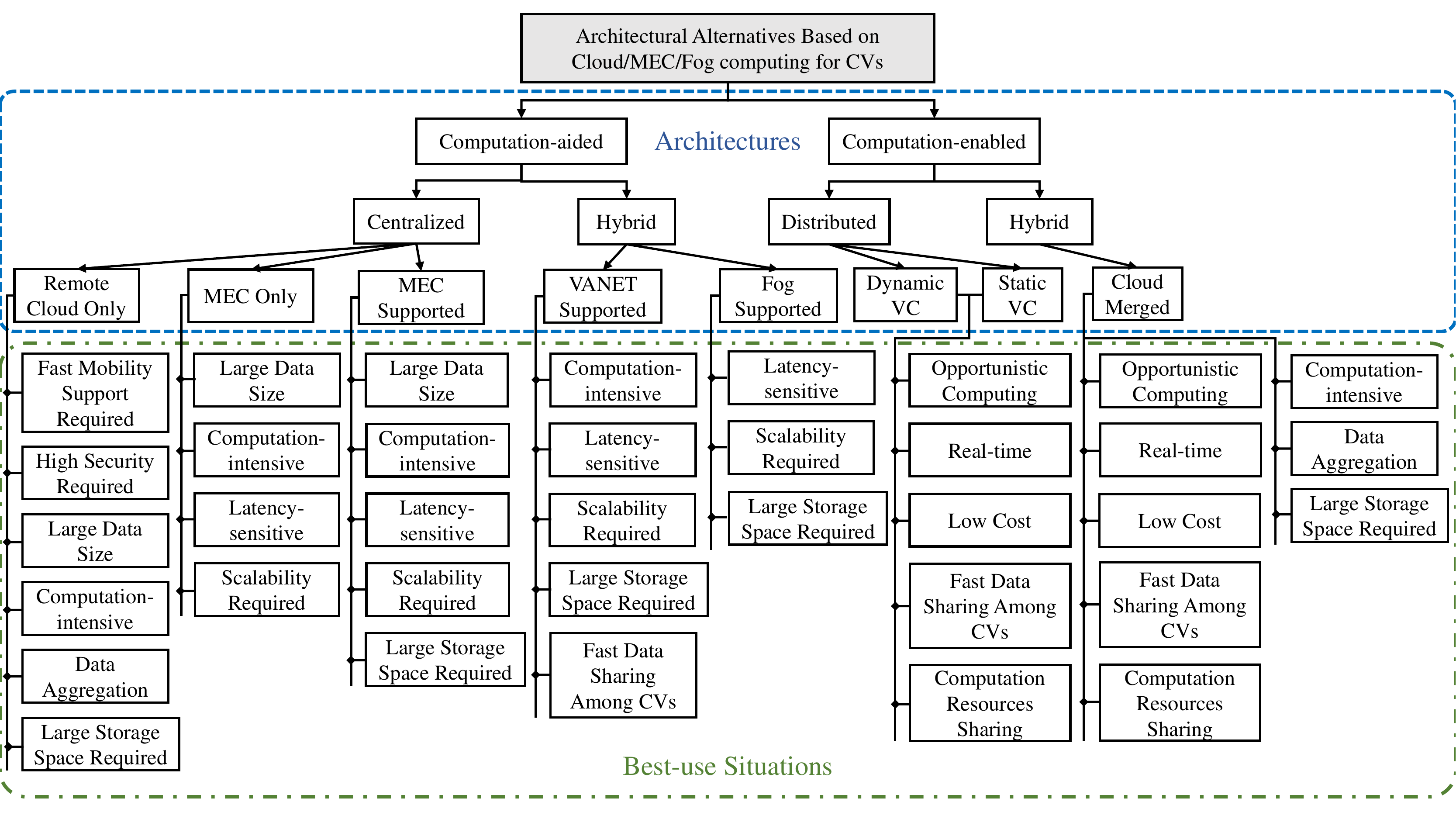}
\caption{Taxonomy of computing architectural design alternatives for CVs.}
\label{Figure2}
\end{figure*}

\begin{table*}[tp!]
\newcommand{\tabincell}[2]{\begin{tabular}{@{}#1@{}}#2\end{tabular}}
\centering
\caption{Design Considerations in Existing Computation-aided Computing Architectures for CVs}
\begin{tabular}{|c|l|c|l|}
\hline
\textbf{Sub-Category} & \textbf{Design Consideration} & \textbf{Reference} & \textbf{Proposed Solution}\\\hline

\multirow{30}{*}{Centralized} & \multirow{4}{*}{Data} & \cite{Wang_2011_WCSP} & \tabincell{l}{Temperature, pressure, image, biomedical information, driver's physical information,\\ GPS, etc.}\\\cline{3-4}
&  & \cite{Li_2011_IIS} & \tabincell{l}{Traffic data and the state of urban transportation.}\\\cline{3-4}
&  & \cite{Wang_2017_TVT} & \tabincell{l}{Air quality data.}\\\cline{2-4}

& \multirow{3}{*}{Application} & \cite{Wang_2011_WCSP} & \tabincell{l}{(i) Context-based services, such as driver status monitoring, road pollution monitoring,\\ vehicle performance monitoring, etc.\\ (ii) Communication-based services, such as road traffic monitoring, weather information,\\ Internet access, etc.\\ (iii) Customized services, such as parking, health-care, resource sharing, etc.}\\\cline{3-4}
&  & \cite{Li_2011_IIS} & \tabincell{l}{Intelligent transportation services such as traffic management.}\\\cline{3-4}
&  & \cite{Wang_2017_TVT} & \tabincell{l}{Metropolitan air quality monitoring.}\\\cline{2-4}

& \multirow{9}{*}{Communication} & \cite{Wang_2017_TVT} & \tabincell{l}{An efficient data gathering and estimation mechanism is proposed to reduce the data\\ transmission overhead while keep monitoring accuracy by dynamically adjust data\\ sampling rate.}\\\cline{3-4}
& & \cite{Yu_2016_ITVT} & \tabincell{l}{5G-Enhanced cloud radio access network is proposed for communication enhancement in\\ terms of throughput, delay, reliability, scalability, and mobility.}\\\cline{3-4}
&  & \cite{Zhang_2017_IVTM} & \tabincell{l}{A computation offloading scheme with predictive-mode transmission is proposed to reduce\\ the data transmission overhead.}\\\cline{3-4}
&  & \cite{Hung_2015_IA} & \tabincell{l}{An approach that jointly optimize the radio resource allocation, allocation of APs in mobile\\ network, and handover together is proposed for vehicular networks.}\\\cline{3-4}
&  & \cite{Malandrino_2014_MC} & \tabincell{l}{A vehicle mobility prediction-based approach is proposed to schedule the data offloading\\ and improve the transmission efficiency. }\\\cline{2-4}

& \multirow{4}{*}{Computation \& Storage} & \cite{Yu_2016_ITVT} & \tabincell{l}{A matrix game theoretical approach is proposed to solve the resource sharing and\\ allocation problems among cloudlets.}\\\cline{3-4}
&  & \cite{Li_2017_ICFEC} & \tabincell{l}{Two local computation resource management schemes, namely fog resource reservation\\ and resource reallocation, are proposed to improve the QoS of latency-sensitive connected\\ services.}\\\cline{2-4}

& \multirow{2}{*}{Security} & \cite{Wang_2011_WCSP} & \tabincell{l}{Security issues such as vehicle driver's privacy protection and V2I authorization and\\ authentication are proposed.}\\\cline{3-4}
&  & \cite{Li_2011_IIS} & \tabincell{l}{VM is utilized to protect users' data and equipment safety.}\\\hline

\multirow{20}{*}{Hybrid} & \multirow{3}{*}{Data}  & \cite{Liu_2017_ICM_zeng,Liu_2017_ICM_jia} & \tabincell{l}{Traffic data and the state of urban transportation.}\\\cline{3-4}
&  & \cite{wan_2014} & \tabincell{l}{Entertainment data and traffic accident information.}\\\cline{3-4}
&  & \cite{Gerla_2013_ICNC} & \tabincell{l}{On-board camera captured images.}\\\cline{2-4}

& \multirow{3}{*}{Application} & \cite{Liu_2017_ICM_zeng,Liu_2017_ICM_jia} & \tabincell{l}{Urban traffic management such as rapid road accident rescue.}\\\cline{3-4}
&  & \cite{wan_2014} & \tabincell{l}{Data sharing services such as entertainment resources and traffic accident information.}\\\cline{3-4}
&  & \cite{Gerla_2013_ICNC}& \tabincell{l}{Photo surveillance service.}\\\cline{2-4}

& \multirow{3}{*}{Communication} & \cite{Liu_2017_ICM_zeng,Liu_2017_ICM_jia} & \tabincell{l}{5G and SDN are introduced into the system architecture to improve the agility, reliability,\\ scalability, and latency performance. }\\\cline{3-4}
&  & \cite{Huang_2018_Access} & \tabincell{l}{A centralized V2V path selection algorithm
is proposed to find the V2V routing path that\\ has the longest life time based on the current network topology.}\\\cline{2-4}

& \multirow{1}{*}{Computation \& Storage} & \cite{zhang_2017} & \tabincell{l}{A flexible hierarchical resource management methodology that includes Intra-fog and\\ Inter-fog resource management is proposed for reducing the system maintenance cost\\ and improving the QoS in areas with high population density.}\\\cline{2-4}

& \multirow{1}{*}{Interaction} & \cite{Liu_2017_ICM_zeng} & \tabincell{l}{MEC to remote cloud server: all traffic data is pre-processed by the MEC server and\\ then delivered to the remote cloud server for further processing such as traffic prediction.}\\\cline{2-4}

& \multirow{7}{*}{Security} & \cite{Huang_2017_IEEA,li2014joint,li2015art} & \tabincell{l}{Distributed reputation management approaches are proposed to avoid malicious attacks and\\ evaluate trustworthiness of vehicles in VANETs.}\\\cline{3-4}
&  & \cite{hasbullah2010denial} & \tabincell{l}{Multiple solutions for the DoS attack in VANETs is discussed.}\\\cline{3-4}
&  & \cite{Soleymani_2017_IA} & \tabincell{l}{A fuzzy trust model is proposed to secure vehicles to receive correct and credible\\ information from surrounding vehicles.}\\\cline{3-4}
&  & \cite{hao2010secure} & \tabincell{l}{A secure data downloading protocol with privacy preservation for VANETs is proposed.}\\\cline{3-4}
&  & \cite{garg2019edge} & \tabincell{l}{A smart security framework for VANETs equipped with edge computing nodes is\\ proposed to provide secure V2V and V2I communications by using the Quotient filter.}\\\hline

\end{tabular}
\label{Table8}
\end{table*}

\begin{table*}[tp!]
\newcommand{\tabincell}[2]{\begin{tabular}{@{}#1@{}}#2\end{tabular}}
\centering
\caption{Design Considerations in Existing Computation-enabled Computing Architectures for CVs}
\begin{tabular}{|c|l|c|l|}
\hline
\textbf{Sub-Category} & \textbf{Design Consideration} & \textbf{Reference} & \textbf{Proposed Solution}\\\hline

\multirow{30}{*}{Distributed} & \multirow{6}{*}{Application} & \cite{eltoweissy2010towards} & \tabincell{l}{Dynamic traffic light management, self-organized high occupancy vehicle (HOV) lanes,\\ and disaster management.}\\\cline{3-4}
&  & \cite{arif2012datacenter} & \tabincell{l}{Airport as a datacenter, where CVs parked in an international airport are used as a basis\\ for a datacenter at the airport.}\\\cline{3-4}
&  & \cite{Feng_2017_TVE} & \tabincell{l}{(i) High-priority applications, such as navigation, routing and information services,\\ optional safety-enhancing applications, etc. (ii) Low-priority applications. such as video\\ streaming, video processing, passenger entertainment, etc.}\\\cline{2-4}

& \multirow{8}{*}{Formation} & \cite{Feng_2017_TVE} & \tabincell{l}{A beacon broadcasting mechanism is proposed for CVs to acquire computation\\ resources nearby.}\\\cline{3-4}
&  & \cite{Abuelela_2011} & \tabincell{l}{A VC formation workflow is proposed which includes broker election, authorization\\ from high authority, inviting neighbor CVs, VC formation time announcement, and\\ resource pooling.}\\\cline{3-4}
&  & \cite{Wang_2017_ITS} & \tabincell{l}{(i) A clustering mechanism is proposed based on the derivation of the optimal cluster\\ length and the design of a time division policy. (ii) A cluster association mechanism is\\ developed for CVs to dynamically join or leave a VC.}\\\cline{3-4}
&  & \cite{Gong_2016_CC,arif2012datacenter} & \tabincell{l}{A stationary VC formation approach is discussed.}\\\cline{2-4}

& \multirow{7}{*}{Communication} & \cite{Feng_2017_TVE} & \tabincell{l}{A job scheduling mechanism is proposed for CVs to decrease the job transmission latency\\ and improve the utility in VC scenarios.}\\\cline{3-4}
&  & \cite{Wang_2017_ITS} & \tabincell{l}{A cooperative data transmission scheduling mechanism is proposed for V2V\\ communication in bidirectional road scenarios.}\\\cline{3-4}
&  & \cite{zhang2008vc} & \tabincell{l}{A MAC layer protocol named VC-MAC is proposed to improve the network throughput\\ and decrease the channel collision.}\\\cline{3-4}
&  & \cite{Yu_2012_MCC} & \tabincell{l}{An adaptive hybrid content routing approach is proposed, which combines both proactive\\ and reactive routing.}\\\cline{2-4}

& \multirow{3}{*}{Computation \& Storage} & \cite{arif2012datacenter} & \tabincell{l}{A resource manager is developed for discovering and managing the dynamically changing\\ VC resources.}\\\cline{3-4}
&  & \cite{kim2018vehicular} & \tabincell{l}{A dynamic resource management strategy is developed based on a checkpoint mechanism\\ by considering individual CV capacity and leave rate.}\\\cline{2-4}

& \multirow{1}{*}{Security} & \cite{whaiduzzaman_2014} & \tabincell{l}{Six groups of threats in VCs are introduced, including DoS, identify spoofing, modification\\ repudiation, repudiation, Sybil attack, and information disclosure.}\\\hline

\multirow{16}{*}{Hybrid} & \multirow{6}{*}{Application} & \cite{Guerrero-ibanez_2015_IWC,Darwish_2018_Access} & \tabincell{l}{Intelligent transportation system.}\\\cline{3-4}
&  & \cite{He_2014_TII} & \tabincell{l}{Intelligent parking cloud service.}\\\cline{3-4}
&  & \cite{Gu_2013_GC_Wkshps}& \tabincell{l}{Autonomous driving management, intermediate cache, participatory sensing, personal\\ cloudlet, content sharing, and traffic management.}\\\cline{3-4}
&  & \cite{Yu_2013}& \tabincell{l}{Real-time navigation with computation resource sharing, video surveillance with storage\\ resource sharing, and cooperative download/upload with bandwidth sharing}\\\cline{2-4}

& \multirow{1}{*}{Communication} & \cite{wang2018computation} & \tabincell{l}{A multi-layer computation offloading architecture is proposed, including the user layer,\\ mobile fog layer, fixed fog layer, and cloud layer; a computation offloading scheme is\\ proposed to maximize the total profits of the offloading from the infrastructure perspective.}\\\cline{2-4}

& \multirow{1}{*}{Computation \& Storage} & \cite{Yu_2013} & \tabincell{l}{A VM resource allocation scheme for VCs and roadside clouds is proposed based on a\\ game-theoretical model.}\\\cline{2-4}

& \multirow{3}{*}{Interaction} & \cite{Zhang_2015_TVT} & \tabincell{l}{A flexible offloading strategy is proposed for performing task migration between VCs and\\ infrastructure-based clouds based on the estimated resource conditions.}\\\cline{3-4}
&  & \cite{Tao_2017_IEEN,Bitam_2015_IWC} & \tabincell{l}{A possible case for the interconnection and interoperation between VCs and infrastructure-\\based clouds is discussed.}\\\cline{2-4}

& \multirow{1}{*}{Security} & \cite{zhou2015secure} & \tabincell{l}{A secure and privacy-preserving packet forwarding scheme is proposed to resist layer\\-adding attacks.}\\\hline

\end{tabular}
\label{Table9}
\end{table*}

\begin{table*}[tp!]
\newcommand{\tabincell}[2]{\begin{tabular}{@{}#1@{}}#2\end{tabular}}
\centering
\caption{The Comparison of Different Cloud/Edge/Fog Computing Architectures for CVs}
\begin{tabular}{|c|c|c|c|c|l|}
\hline
\multirow{2}{*}{\textbf{Category}} & {\textbf{Sub}} & {\textbf{On-Board}}  & {\textbf{External}} & \multirow{2}{*}{\textbf{Reference}} & \multirow{2}{*}{\textbf{Proposed Architecture}}\\
 & \textbf{Category} & \textbf{Computation} & \textbf{Computation} & &\\\hline
 
\multirow{20}{*}{Computation-aided} & \multirow{8}{*}{Centralized} & \multirow{8}{*}{Disabled} & \multirow{8}{*}{Enabled} & \cite{Wang_2011_WCSP} & \tabincell{l}{Propose a three-layer VC architecture that includes\\ device, communication, and service levels, from the\\ perspective of communication support.}\\\cline{5-6}
 &  &  &  & \cite{Li_2011_IIS} & \tabincell{l}{Propose a four-layer agent-based intelligent traffic\\ cloud architecture that includes application, platform,\\ unified source, and fabric layers, from the perspective\\ of computation support.}\\\cline{5-6}
 &  &  &  & \cite{Wang_2017_TVT}&\tabincell{l}{Propose a centralized metropolitan air quality\\ monitoring architecture to mitigate the trade-off between\\ monitoring accuracy and data offloading cost.}\\\cline{5-6}
 
\cline{2-4}& \multirow{12}{*}{Hybrid} & \multirow{12}{*}{Disabled} & \multirow{12}{*}{Enabled} & \cite{Liu_2017_ICM_zeng} & \tabincell{l}{Propose to extend the centralized architecture by\\ utilizing MEC and to improve the network capacity by\\ designing a 5G-enabled vehicular network.} \\\cline{5-6}
&  &  &  & \cite{Liu_2017_ICM_jia} & \tabincell{l}{Propose a four-layer architecture that includes\\ environment sensing, communication, MEC server, and\\ remote core cloud server layers, for urban traffic\\ management with the convergence of $5$G networks,\\ VANETs, MEC, and software defined networks (SDNs)\\ technologies.}\\\cline{5-6}
& &  &  &\cite{wan_2014} & \tabincell{l}{Propose a three-layer hybrid computing architecture\\ that includes micro, meso, and macro layers, where\\ the proposed vehicular cluster only performs the\\ function of data sharing.}\\\cline{5-6}
 &  &  &  & \cite{zhang_2017,Ge_2017_ICM} & \tabincell{l}{Propose a two-layer cooperative fog architecture that\\ includes edge and fog layers.}\\\cline{5-6}
\hline

\multirow{12}{*}{Computation-enabled} & \multirow{4}{*}{Distributed} & \multirow{4}{*}{Enabled} & \multirow{4}{*}{Disabled} & \cite{Feng_2017_TVE,Abuelela_2011,Wang_2017_ITS}& \tabincell{l}{Propose a distributed architecture that provides\\ computation services in dynamic vehicular\\ environments via managing the idle computational\\ resources on each CV.}\\\cline{5-6}
&  &  &  & \cite{Sookhak_2017_VTM,Gong_2016_CC} & \tabincell{l}{Propose a static distributed architecture to augment the\\ computation and storage power of fog computing by\\ utilizing a pool of parked CVs.}\\\cline{5-6}

\cline{2-4}& \multirow{7}{*}{Hybrid} & \multirow{7}{*}{Enabled} & \multirow{7}{*}{Enabled} & \cite{Guerrero-ibanez_2015_IWC,He_2014_TII,Gu_2013_GC_Wkshps,Yu_2013,Zhang_2015_TVT} & \tabincell{l}{Propose a hybrid computing architecture by merging\\ vehicular clouds with cloud computing.}\\\cline{5-6}
&  &  &  & \cite{Tao_2017_IEEN,Bitam_2015_IWC} & \tabincell{l}{Propose a two-layer hybrid architecture that includes\\ permanent and temporary clouds.}\\\cline{5-6}
&  &  &  & \cite{wang2018computation} & \tabincell{l}{Propose a four-layer hybrid architecture that includes\\ user, mobile fog, fixed fog, and cloud layers and an\\ algorithm to maximize the total profits of computation\\ offloading from the infrastructure perspective.}\\\cline{5-6}
\hline

\end{tabular}
\label{Table10}
\end{table*}

In this section, we survey the existing proposed computing architectural alternatives for CVs. They can be broadly classified into two categories: \textit{computation-aided} computing architecture, where CVs only generate computing tasks and do not possess the computation ability, and \textit{computation-enabled} computing architecture, where CVs not only generate computing tasks but obtain computation capabilities. 

In computation-aided computing architectures, external infrastructures (e.g., the cloud server, edge servers, and fog servers) are the only computation resources for CVs that are sources of data. Several non-negligible design considerations for computation-aided computing architectures are briefly described as follows: 
\begin{itemize}
\item \textbf{Data:} What kind of data might be generated or collected from the CVs (e.g., driver status, road traffic, weather information, etc.)?
\item \textbf{Application:} What kind of services or applications are provided by the cloud, edge, and fog servers (e.g., customized CV services, intelligent transportation, surveillance, etc.), which is crucial because different services or applications may have completely different requirements (e.g., latency-sensitive, requiring a large amount of collected data, computation-intensive, etc.)?
\item \textbf{Communication:} How to enable efficient data transmissions between CVs and external infrastructures due to the limited network resources (e.g., deploying advanced communication technologies, enabling V2V or cooperative-relay transmission, deploying smart path selection or routing strategies, etc.)?
\item \textbf{Computation \& Storage:} How to manage the computation and storage resources of the external infrastructures?
\item \textbf{Interaction:} How the cloud, edge, and fog servers interact and coordinate with each other?
\item \textbf{Security:} How can security and privacy be ensured in computation-aided computing architectures?
\end{itemize}

In computation-enabled computing architectures, CVs might be not only the sources of data but also the sources of computation. Besides external infrastructures, the idle computation resources on each CV can be shared with other CVs. Thus, VCC/VFC is a key component in computation-enabled computing architectures, where ``a group of largely autonomous vehicles whose corporate computing, sensing, communication and physical resources can be coordinated and dynamically allocated to authorized users" \cite{whaiduzzaman_2014,Security_Gong_2013}. Several non-negligible design considerations for computation-enabled computing architectures are briefly described as follows: 
\begin{itemize}
    \item \textbf{Data:} What kind of data might be generated or collected from the CVs?
    \item \textbf{Application:} What kind of services or applications are provided by the VCC/VFC and external infrastructures?
    \item \textbf{Formation:} What are the possible VCC/VFC formation scenarios?
    \item \textbf{Communication:} How to enable efficient data transmissions between (i) CVs within the same VCC/VFC, (ii) CVs in different VCCs/VFCs, and (iii) VCC/VFC and external infrastructures?
    \item \textbf{Computation \& Storage:} How to manage the computation and storage resources of the VCC/VFC?
    \item \textbf{Interaction:} How the VCC/VFC interacts and coordinates with the external infrastructures?
    \item  \textbf{Security:} How can security and privacy be ensured in computation-enabled computing architectures?
\end{itemize}
Note that not all of the aforementioned design considerations are taken into account in each existing work. Table \ref{Table8} and \ref{Table9} present brief summaries of the state-of-the-art solutions of design considerations for both computation-aided and -enabled computing architectures.

Under each category, computing architectures can be further divided into centralized, distributed, and hybrid architectures based on the distribution of the computation resources.
In \textit{centralized} computing architectures, computing and storage resources are organized in a remote centralized server. Centralized architectures may also have hierarchical computing where computing and storage resources are organized in a hierarchical structure from the edge of networks to the remote center. In \textit{distributed} computing architectures, computing resources are distributed in a number of individual units without the support of a centralized controller. \textit{Hybrid} computing architectures combine centralized and distributed computing architectures. Fig. \ref{Figure2} and Table \ref{Table10} present a taxonomy and a comparison of computing architectural design alternatives for CVs, respectively, which will be discussed in detail in the following sections.

\subsection{Computation-aided Computing Architectures}\label{sec:Centralized}

The computing architectures considered in papers
\cite{Wang_2011_WCSP,Li_2011_IIS,Wang_2017_TVT,khayyam_2013,Zhang_2016_IEEC,Nunna_2015_ICIT,Zhang_2017_IVTM,Kumar_2016_ICM,Yu_2016_ITVT,He_2017_IEAA,Hung_2015_IA,Li_2017_ICFEC,Malandrino_2014_MC,Stojmenovic_2014_ATNAC,Liu_2017_ICM_jia,Liu_2017_ICM_zeng,wan_2014,Jaworski_2011_ITSC,zhang_2017,Gerla_2013_ICNC,Truong_2015_IM,Deng_2017_ICM,Chaba_2017_Comptelix,Ni_2017_ICM,Huang_2018_Access,Zhao_2018_JSAC,Ge_2017_ICM,8761551,Huang_2017_IEEA,li2014joint,hao2010secure,hasbullah2010denial,Soleymani_2017_IA,li2015art,wang2020user,wang2018smart,wang2019globe} for supporting CVs are computation-aided architectures, where CVs only generate computing tasks (i.e., CVs are considered as computation sources only), leaving task computation to external units, such as nearby edge nodes (e.g., WiFi routers, small-cell BSs, and macro-cell BSs) or cloud servers, which depends on the service requirements on the computation load. Since CVs in computation-aided computing architectures do not possess the computation ability, their generated tasks must be offloaded to external computing infrastructures. Therefore, distributed architectures, under which CVs usually compute their generated tasks utilizing their own on-board or other nearby vehicle clusters' computation resources, will not be discussed in this subsection.

\subsubsection{Centralized Architecture \cite{Wang_2011_WCSP,Li_2011_IIS,Wang_2017_TVT,khayyam_2013,Zhang_2016_IEEC,Nunna_2015_ICIT,Zhang_2017_IVTM,Kumar_2016_ICM,Yu_2016_ITVT,He_2017_IEAA,Hung_2015_IA,Li_2017_ICFEC,Malandrino_2014_MC,Stojmenovic_2014_ATNAC}}
\begin{figure}[tp!]
\centering
\includegraphics[width=0.48\textwidth]{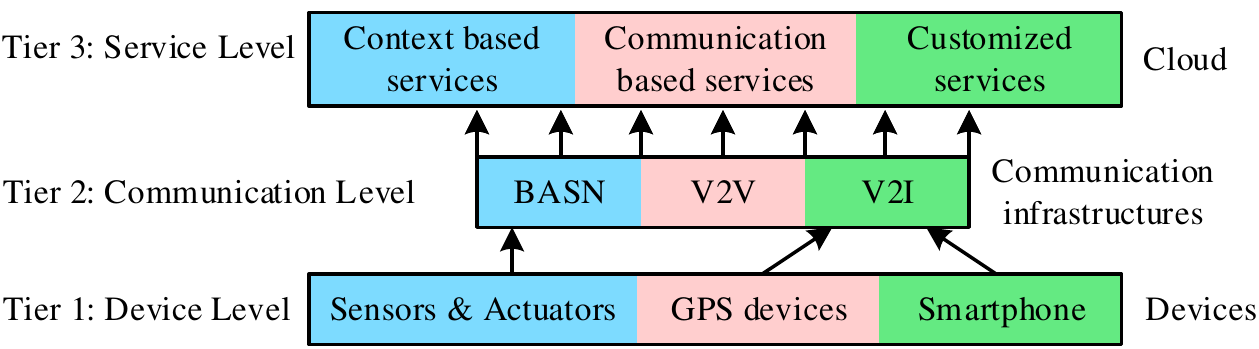}
\caption{The proposed three-tier vehicle cloud architecture in paper \cite{Wang_2011_WCSP}.}
\label{Figure3}
\end{figure}

Paper \cite{Wang_2011_WCSP} proposed a three-layer VC architecture from the perspective of communication support. This architecture is composed of the device level, communication level, and service level, as shown in Fig. \ref{Figure3}. At the device level, various devices ranging from sensors, actuators, Global Positioning System (GPS) devices, and smartphones are used for collecting data such as temperature, pressure, image, and driver's bio-medical information. Then, these collected raw data are stored in a repository and wait for further processing at the upper level. Based on the pre-processing techniques, those stored raw data can be classified into high-level context (such as human activity and gesture) and low-level context (such as pressure and temperature). The communication level is divided into in-car communication modules (e.g., BASNs), V2V communication modules, and V2I communication modules (e.g., satellite and cellular networks).  The service level includes various services such as context-based services (e.g., driver status monitoring), communication-based services (e.g., road traffic monitoring, weather information, and Internet access), and customized services (e.g., parking, health-care, and dining booking). Context-based services are given charge of tasks that include drivers' health and safety improvements, while tasks like drivers' convenience and comfort degree improvements are allocated to communication-based services. As stated above, this proposed three-tier architecture collects a wide variety of data on the device level. Thus, customized CV services that require various data and high accuracy are suitable for being executed in this architecture.

\begin{figure}[tp!]
\centering
\includegraphics[width=0.48\textwidth]{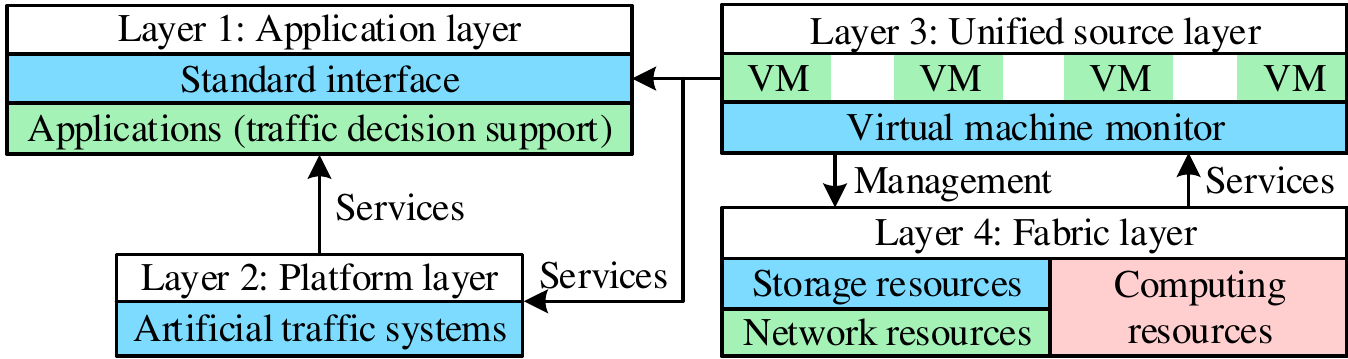}
\caption{The proposed four-layer intelligent transportation cloud architecture in paper \cite{Li_2011_IIS}.}
\label{Figure4}
\end{figure}

In paper \cite{Li_2011_IIS}, a four-layer agent-based intelligent traffic cloud architecture from the perspective of computation support is proposed. This proposed architecture mainly focuses on the scenario of intelligent transportation systems and aims to handle a large amount of computing and storage resources that are required to use traffic strategy agents and massive transport data. It includes the application layer, platform layer, unified source layer, and fabric layer. The application layer shown in Fig. \ref{Figure4}, contains all the applications that run in the clouds, including agent management, generation, optimization, testing, traffic decision support, and agent-oriented task decomposition. Customers can obtain the services that they need through a pre-defined standard interface. The platform layer is made of artificial transportation systems, providing platform as a service. Components such as weather simulator, population synthesizer, $3$D game engine, and path planner are contained in this layer and provide services to the traffic applications in the upper application layer. The unified source layer maintains the hardware resources in the lower fabric layer and provides infrastructure as a service. In the unified source layer, VMs are utilized to protect users' data and equipment safety. In addition, a unified access interface is established for the upper distribute computing resources. These features described above can help the intelligent transportation system efficiently mine useful knowledge from the massive urban traffic data. Furthermore, hardware-level resources (e.g., storage, network, and computing resources) are contained in the fabric layer, which will help the intelligent traffic cloud supply the peak demand of urban-traffic management systems. CV services with requirements such as large data size, high security, multiple users, latency-insensitive, and high computing power are recommended for using this four-layer architecture.

As stated above, one of the advantages of centralized cloud architectures is data aggregation. By using the cloud storage techniques, the cloud can provide a variety of stored data for the private and government agencies (e.g., department of transportation, the meteorology department) to perform various studies. Therefore, several papers described various studies in vehicular systems based on centralized cloud computing architectures. For example, paper \cite{Wang_2017_TVT} proposed a metropolitan air quality monitoring service. In this service, vehicles act as the air quality data gathering sensors under a similar architecture described above. Vehicles offload their gathered data to a remote centralized cloud for performing an air quality estimation. The main contribution of this paper is the investigation of the trade-off between monitoring accuracy and data offloading overhead, where dynamic grid partition and probabilistic reporting approaches are proposed to adjust data sampling rate and avoid redundant data.

However, with the development of the latency-critical and computation-intensive applications in CVs, centralized architectures with a remote cloud server described above are not efficient. Therefore, architectures of CV networks with MEC are proposed, where besides a centralized cloud server, additional computing infrastructures (e.g., RSUs, BSs, and/or edge/fog servers) are also given the role of computing units, forming a hierarchical computing architecture. The main objective of MEC is to extend the cloud computing functionality to the edge of networks, which saves network bandwidth and reduces the communication latency. For example, several papers \cite{Nunna_2015_ICIT,Yu_2016_ITVT,Stojmenovic_2014_ATNAC} proposed to utilize 5G and MEC together to help CVs improve task transmission efficiency. Among these papers, paper \cite{Yu_2016_ITVT} proposed a paradigm of $5$G-enabled vehicular networks to improve the network capacity. The cloud server is extended by integrating geographically distributed cloudlets that are responsible for local services. In addition, the matrix game theoretical approach is exploited to operate the resource sharing and allocation among cloudlets. 

\begin{figure}[tp!]
\centering
\includegraphics[width=0.48\textwidth]{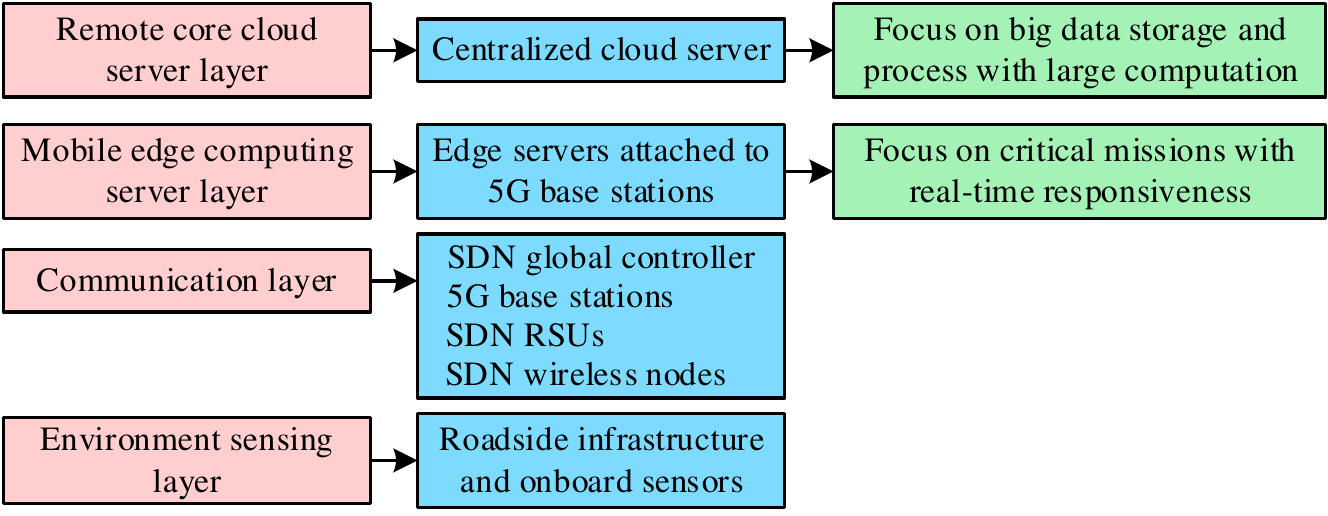}
\caption{The proposed 4-tier hierarchical computing architecture in papers \cite{Liu_2017_ICM_jia,Liu_2017_ICM_zeng}.}
\label{Figure5}
\end{figure}

In addition, other variations of hierarchical computing architectures for CVs were also proposed. Paper \cite{Zhang_2017_IVTM} proposed a cloud-based MEC offloading framework in vehicular networks. Considering the time consumption of the computation task execution and the mobility of the vehicles, data are adaptively offloaded to the MEC servers through a direct uploading mode, i.e., V2I, or a predictive relay mode, i.e., V2V. Furthermore, the proposed framework does not have a remote core cloud server and all computing tasks are offloaded to different MEC servers, which reduces the transmission cost and the latency of the computation offloading. However, since MEC servers are deployed at the edge of networks, e.g., RSUs, their computation capacities, storage, and service ranges are limited. Thus, applications require high computation resource, a large amount of feed data, and constant offloading environment with fast mobility may be seriously restricted by this proposed framework. In papers \cite{Hung_2015_IA,Li_2017_ICFEC}, fog servers are proposed to co-locate with BSs, forming a BS-fog node. However, all of these proposed architectures face a common challenge -- mobility management issue. In order to mitigate the impact of mobility, paper \cite{Malandrino_2014_MC} proposed a mobility prediction mechanism where RSUs exploit mobility predictions to decide which data they should fetch from the Internet and to schedule the further transmission to vehicles.

\subsubsection{Hybrid Architecture \cite{Liu_2017_ICM_jia,Liu_2017_ICM_zeng,wan_2014,Jaworski_2011_ITSC,zhang_2017,Gerla_2013_ICNC,Truong_2015_IM,Deng_2017_ICM,Chaba_2017_Comptelix,Ni_2017_ICM,Huang_2018_Access,Zhao_2018_JSAC,Ge_2017_ICM,8761551,Huang_2017_IEEA,li2014joint,hao2010secure,hasbullah2010denial,Soleymani_2017_IA,li2015art}}

Papers \cite{Liu_2017_ICM_jia,Liu_2017_ICM_zeng} proposed a four-layer architecture for urban traffic management with the convergence of $5$G networks, VANETs, MEC, and SDNs technologies, as shown in Fig. \ref{Figure5}. It contains the environment sensing layer (e.g., traffic data are derived from the roadside infrastructure and on-board sensors), communication layer, including SDN global controller, 5G BSs, SDN RSUs, and SDN wireless nodes (e.g., vehicles), MEC server layer (MEC servers are deployed in 5G BSs), and remote core cloud server layer. The communication layer with two emerging network paradigms, 5G and SDN, provides several advantages. First, since the data and control planes are separated, forwarding policies can be exploited to balance the traffic flows, and provide a more flexible path selection strategy with the network's programmability. Second, in order to support different approaches of communication, several wireless modules are deployed on vehicles. Thus, the SDN-based communication layer can provide a path selection strategy based on the cognitive radio and channel allocation policy, which may reduce the communication latency and increase the communication bandwidth in CV networks. Thus, the channel and frequency selection in CV networks can be more flexible. Third, the 5G cellular network with multiple-input multiple-output (MIMO), wide spectrum, and ultra-dense network technologies can achieve $1.2$ Gb/s data rate in a mobile environment, where a vehicle is at the speed of $100$ km/h, based on 28 GHz spectrum \cite{Liu_2017_ICM_jia}. Fourth, because the current cellular network is designed for mobile broadband traffic, it lacks support for V2V communications. Although IEEE 802.11p is proposed for V2V communications, its limited bandwidth and peak data rate (i.e., up to 27 Mbps \cite{cunha2016data,lu2014connected}) might not satisfy 
the diverse requirements of vehicular applications \cite{shah20185g}. The 5G Public Private Partnership (5G-PPP) proposed device-to-device (D2D) communications that data can be directly exchanged among mobile users by bypassing infrastructure within 1 ms delay \cite{lee2016lte}, which demonstrates that 5G is a promising approach for V2V communications. In addition, paper \cite{shah20185g} presents three salient features of 5G-enabled communications in vehicular scenarios, including proximity service, integration of MEC, and network slicing. For example, (i) proximity service provides a solid foundation for vehicular safety communications and identifies the source of autonomous vehicle attacks; (ii) MEC plays a fundamental role in 5G \cite{mao2017survey}, which can improve the user experience of vehicular applications such as the traffic information system that have flexible latency requirements; and (iii) variant network slices can be designated based on the diverse requirements of vehicular applications, which simplifies the design of vehicular systems.
Compared with the centralized cloud computing architecture, mobile edge servers are deployed closer to end users in this architecture. Thus, it reduces the delay of data offloading and is suitable for implementing latency-sensitive applications. A rapid road accident rescue system, for instance, can be built under this architecture with the support of the low-latency and high-bandwidth SDN-based heterogeneous network and the fast-response MEC server.

\begin{figure}[tp!]
\centering
\includegraphics[width=0.48\textwidth]{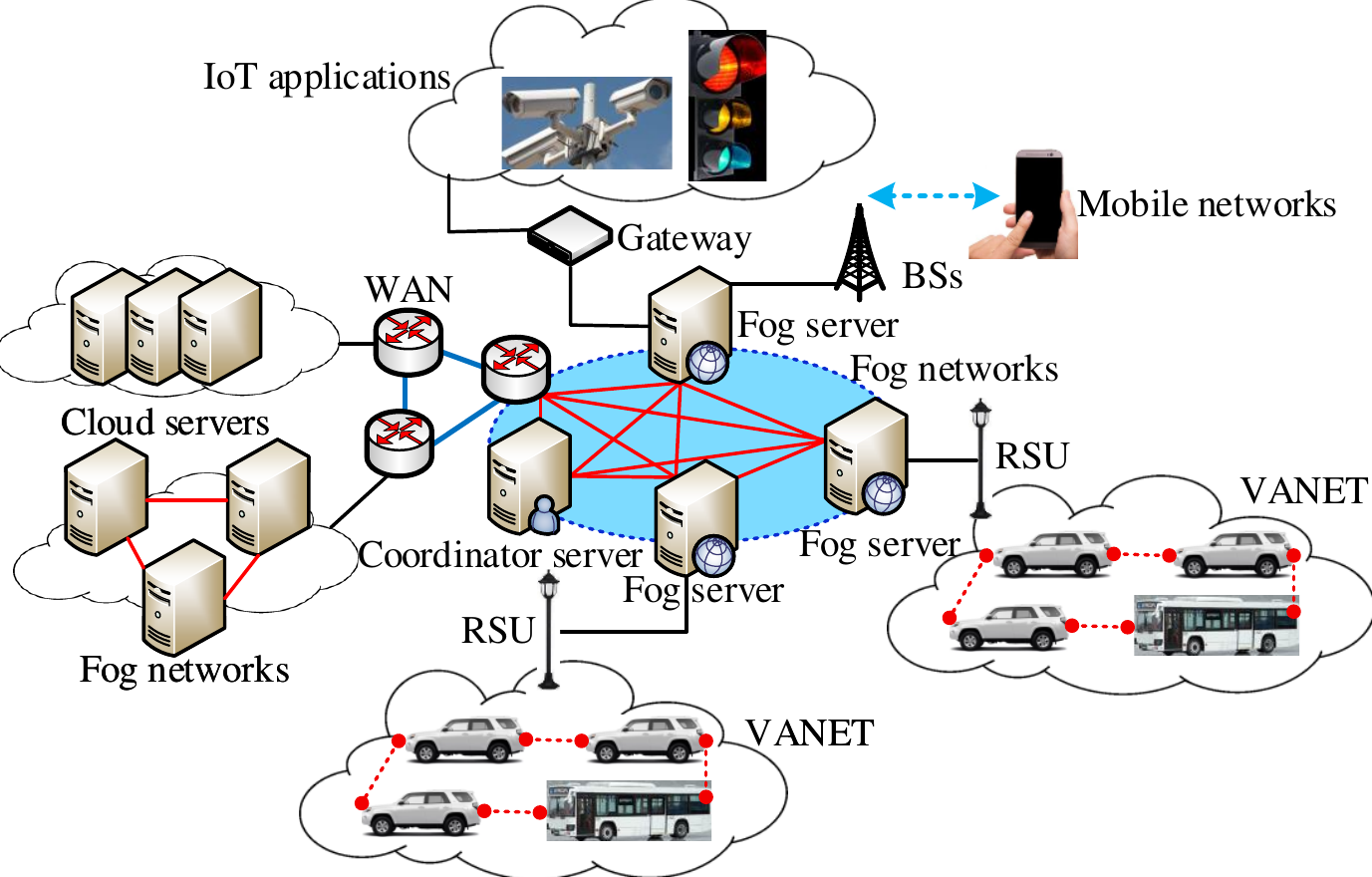}
\caption{The proposed hybrid vehicle cloud architecture in paper \cite{zhang_2017}.}
\label{Figure6}
\end{figure}

Paper \cite{wan_2014} proposed a three-layer hybrid computing architecture, including the micro layer (vehicles and users), meso layer (transmission), and macro layer (cloud services). However, the vehicular cluster under this architecture only performs the function of sharing data, e.g., entertainment resources and traffic accident information among vehicles in this cluster, instead of providing computing services.
Papers \cite{zhang_2017,Ge_2017_ICM} introduced a new concept, fog computing, into the cloud computing architecture for CVs, where a cooperative fog architecture, shown in Fig. \ref{Figure6}, is proposed. The cooperative fog architecture mainly contains two layers: edge layer and fog layer. The edge layer may include the components such as  VANETs (i.e., a VANET can be applied for V2V communications and traffic information broadcasting), IoT (i.e., lots of IoT application are widely used in city transportation systems such as video traffic surveillance systems, range finders, and wireless sensors), and mobile cellular networks. The fog layer is a federation of geographically distributed local fog servers and may include entities such as fog servers (i.e., the long data transmission latency between CVs and the cloud server can be reduced via offloading the computation and data in the edge layer to local fog servers), access control routers (i.e., they are responsible for controlling or migrating the input data flow), cloud server (i.e., it is deployed out of the fog layer and has strong computation and storage capacity), and coordinator server (i.e., it is responsible for the federation and autonomy of fog networks). Four potential functions might be achieved in this architecture: mobility control, multi-source data acquisition, distributed computation and storage, and multi-path data transmission. Additionally, a flexible hierarchical resource management methodology that includes Intra-fog and Inter-fog resource management is designed for reducing the system maintenance cost and improving the QoS in areas with high population density. Paper \cite{Huang_2018_Access} proposed a hybrid architecture that offloads the vehicular communication traffic in cellular networks to V2V paths based on the SDN. A centralized V2V path selection approach, a lifetime-based network state routing algorithm, is developed based on the SDN inside the MEC architecture, where each CV reports its location, speed, direction, and IDs of the neighboring vehicles to a context database implemented in the MEC server. The proposed approach can not only find the V2V routing path that has the longest life time but also recover a broken V2V.

Several papers proposed multiple CV services in hybrid architectures. For instance, paper \cite{Gerla_2013_ICNC} described a photo surveillance service, named Pics-on-Wheels, that a group of vehicles in a certain area are selected to take camera images of a required urban landscape based on the requirements of a customer. CVs that participate in this service should periodically offload their GPS location to the cloud manager. The customer who requests the service has to send a message containing the time and location to the centralized cloud manager first. Then, the cloud manager will search for available vehicles in the requested time and location by using the best vehicle selection algorithm proposed in this paper. This service can assist with forensic purpose where an accident occurred.

The security and privacy issues in computation-aided hybrid architectures can be divided into two categories, VANET and V2I. As we presented, the VANET is one of key components of computation-aided hybrid architectures. Thus, it is crucial to meet the critical security requirements of VANETs for designing computation-aided hybrid architectures. Firstly, one of the most prevalent security issues is how to maintain the availability of each V2V connection in VANETs, where Denial of Service (DoS) is considered as one of the potential attacks that may affect VANETs \cite{elsadig2016vanets}. Several possible solutions for the DoS attack is discussed in paper \cite{hasbullah2010denial}. Secondly, in order to secure the integrity and ensure the reliability of applications trust must be developed among vehicles in VANETs. In paper \cite{Soleymani_2017_IA}, a fuzzy trust model is proposed to secure vehicles to receive correct and credible information from surrounding vehicles. A series of security checks is conduct by the proposed trust model to ensure the correctness of the received information. Lastly, papers \cite{Huang_2017_IEEA,li2014joint,li2015art,lin2007gsis} jointly investigates the reputation management and privacy protection in VANETs, where reputation management is responsible for rewarding the complying vehicles and punishing the misbehaving ones. A joint privacy and reputation assurance scheme is proposed to reconcile the requirement conflicts of the privacy protection and the reputation management in VANETs. In paper \cite{Huang_2017_IEEA}, edge servers are adopted to perform local reputation management tasks for vehicles, while, in papers \cite{li2014joint,li2015art}, vehicles' reputation values are updated by themselves. Thus, the solution proposed in paper \cite{Huang_2017_IEEA} is said to provide more reliable reputation manifestation and more accurate reputation update than the other two solutions.

Furthermore, V2I is also considered as an important component of computation-aided hybrid architectures, where V2I communication enables vehicles to access services from the cloud/edge/fog computing infrastructures. Preservation of the confidentiality in the V2I communication is one of the most important security requirements. Paper \cite{kim2010efficient} proposes a secure and efficient mutual authentication and key agreement scheme for V2I communications to defend against the RSU replication attack and to prevent all entities from eavesdropping. Paper \cite{hao2010secure} investigates security challenges in wireless communications between CVs and RSUs and proposed a protocol which enables CVs to download data securely from RSU with privacy preservation in VANETs. In addition to that, a smart security framework for VANETs equipped with edge computing nodes is proposed in paper \cite{garg2019edge} to provide secure V2V and V2I communications by using the Quotient filter which is a probabilistic data structure.

\subsection{Computation-enabled Computing Architectures}\label{sec:Distributed}

The architectures considered in papers \cite{Feng_2017_TVE,Yu_2012_MCC,Meneguette_2017_ICC,Wang_2017_ITS,Sookhak_2017_VTM,Gerla_2014_WF-IoT,Gong_2016_CC,Li_2017_IA,Wang_2016_ITITS,Datta_2017_IVTM,Lee_2014_ICM,Hou_2016_ITVT,Abuelela_2011,stojmenovic_2014,whaiduzzaman_2014,Gerla_2012_Med-Hoc-Net,Cui_2018_TIT,He_2014_TII,Gu_2013_GC_Wkshps,Hussain_2012_IC,Yu_2013,huang2017vehicular,Xiao_2017_IIC,Zhang_2015_TVT,Q_2018_IN_Toward,Soua_2018_ICIN,Khan_2018_CCNC,Tao_2017_IEEN,Bitam_2015_IWC,zhou2015secure,arif2012datacenter,eltoweissy2010towards,kim2018vehicular} for supporting CVs are computation-enabled architectures, where CVs not only generate computing tasks, but have computation capabilities. In other words, tasks generated by CVs will be computed by themselves, other nearby vehicle clusters, or with the cooperation of remote clouds. Thus, centralized architectures, under which CVs offload their tasks directly to the remote cloud, will not be discussed in this subsection.

\subsubsection{Distributed Architecture \cite{Feng_2017_TVE,Yu_2012_MCC,Meneguette_2017_ICC,Wang_2017_ITS,Sookhak_2017_VTM,Gerla_2014_WF-IoT,Gong_2016_CC,Li_2017_IA,Wang_2016_ITITS,Datta_2017_IVTM,Lee_2014_ICM,Hou_2016_ITVT,Abuelela_2011,stojmenovic_2014,whaiduzzaman_2014,Gerla_2012_Med-Hoc-Net,Cui_2018_TIT,arif2012datacenter,eltoweissy2010towards,kim2018vehicular}}
VC is new technological shifting, where a cluster of vehicles are in the role of corporate computing, sensing, communication, and data sharing units. In other words, ``a group of CVs whose physical resources can be coordinated and dynamically allocated to authorized users" \cite{eltoweissy2010towards}. Since every CV can be a computing unit, we consider VC as distributed computing. There are three major objectives of VC: (i) it provides low-cost computational services to the authorized users (e.g., vehicle drivers); (ii) it helps minimize road traffic congestion, travel time and accidents; (iii) it offers real-time and low energy consumption services of software, platform, and infrastructure with QoS to drivers. Architectures of VC can be classified into two categories, \textit{dynamic} VC and \textit{static} VC.

Regarding dynamic VC, papers \cite{Feng_2017_TVE,Wang_2017_ITS,Datta_2017_IVTM,Lee_2014_ICM,Abuelela_2011} proposed a distributed computing architecture that provides computation services in dynamic vehicular environments via managing the idle computational resources on each vehicle and utilizing them efficiently. This proposed architecture contains three types of vehicles named requesters, processors, and forwarders. Vehicles that generate jobs are requesters, while others that are responsible for processing these jobs are processors. Forwarders are responsible for relaying the generated jobs to nearby available processors. In addition, a vehicle may be a processor for one job and also be a requester for another job that it does not have sufficient capacity to process. Paper \cite{Yu_2012_MCC} investigated a VC service, content-based routing, that allows VC applications to store, share, and search data within the cloud. Regarding static VC, papers \cite{Sookhak_2017_VTM,Gong_2016_CC,arif2012datacenter,kim2018vehicular} proposed a static architecture of VC to augment the computation and storage power of fog computing. Under this static VC architecture, a pool of smart vehicles parked at a shopping mall, or parked vehicles on the roadside are composed of a computing cloud. In addition, paper \cite{Meneguette_2017_ICC} investigated the service migration issue among different VCs. 

However, the network capacity is considerably limited in a vehicular environment, which may significantly constraint the data sharing and cooperations among vehicles in VC scenarios. In paper \cite{Feng_2017_TVE}, a VC framework that focuses on processor discovery and job scheduling is proposed to decrease the job transmission and processing latency and improve the total utility. The proposed framework consists of three submodules: a job queue module, a resource management module, and a scheduling module. The job queue module is responsible for caching jobs in a CV, which avoids channel contention caused by multiple simultaneous job transmissions. The resource management module controls the available on-board computation resources of a CV, while the scheduling module is responsible for communicating with other CVs, determining the job assignment, offloading jobs and receiving feedbacks. In addition to that, a MAC layer protocol named Vehicular Cooperative Media Access Control is proposed in paper \cite{zhang2008vc} to improve the network throughput and decrease the channel collision. 

Compared to security systems in traditional clouds that are not associated with vehicles, security systems in VCs face more complicated challenges. In the VC scenario, it is difficult to locate an attacker because it is physically moving with a high speed, which may cause several security issues, such as secure location and localization, authentication, data security, and VC access control \cite{whaiduzzaman_2014}. In addition to that, attackers in VCs can pretend to be both computation providers and requesters, which increases the complexity of designing a secure scheme to identify the attackers in VCs. Furthermore, security schemes for VCs must be capable of overcoming a dynamically changing number of vehicles \cite{Security_Gong_2013}, where the number of vehicles in a CV may dynamically change due to the traffic volume, time, terrain, etc. Lastly, security issues in VCs includes security of both networks (e.g., VANET, V2V, V2I, etc.) and cloud computing. Although some security approaches for vehicular networks are applicable in VCs, few specific solutions are developed for VCs. Therefore, in order to make the VC become reality, the challenges of assuring trust and security in VCs need to be addressed \cite{eltoweissy2010towards}. However, to the best of our knowledge, few of existing work investigate the security challenges presented above.

\subsubsection{Hybrid Architecture \cite{Guerrero-ibanez_2015_IWC,He_2014_TII,Gu_2013_GC_Wkshps,Hussain_2012_IC,Yu_2013,huang2017vehicular,Xiao_2017_IIC,Zhang_2015_TVT,Q_2018_IN_Toward,Soua_2018_ICIN,Khan_2018_CCNC,Tao_2017_IEEN,wang2018computation,Bitam_2015_IWC,Darwish_2018_Access,zhou2015secure}}

\begin{figure}[tp!]
\centering
\includegraphics[width=0.42\textwidth]{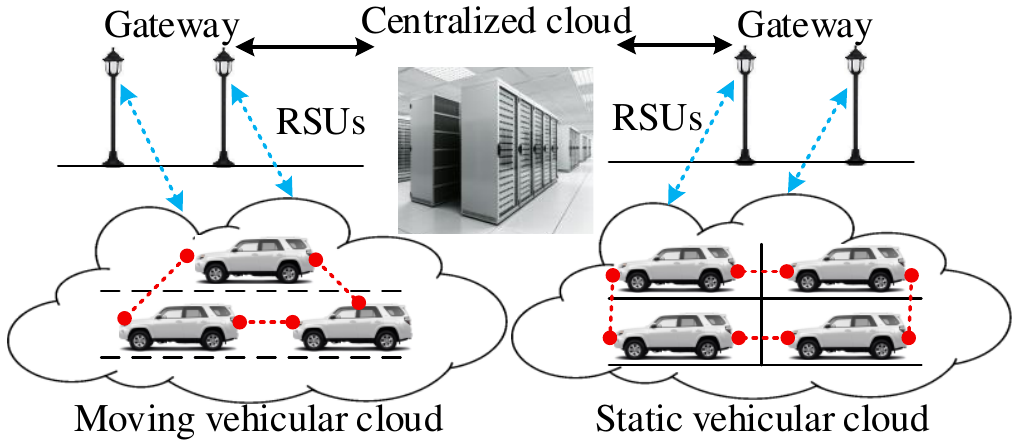}
\caption{The proposed hybrid vehicle cloud architecture in papers \cite{Gu_2013_GC_Wkshps,Hussain_2012_IC,Yu_2013,Zhang_2015_TVT}.}
\label{Figure7}
\end{figure}

Papers \cite{Guerrero-ibanez_2015_IWC,He_2014_TII,Gu_2013_GC_Wkshps,Hussain_2012_IC,Yu_2013,Zhang_2015_TVT, wang2018computation,Soua_2018_ICIN,zhou2015secure} propose to merge VCs with cloud computing to form a hybrid computing architecture, as shown in Fig. \ref{Figure7}, where RSUs act as gateways for VCs to access the centralized cloud. High-speed wired communications can be used for connecting RSUs with the centralized cloud. VCs are further divided into two cases, moving VCs (i.e., a cluster of vehicles on the road) and static VCs (i.e., a cluster of vehicles in a parking lot). For example, some vehicles may need specific applications that require a large number of computing resources or storage space. Therefore, vehicles that have unused storage space can share their computing resources or storage space as a cloud-based service. In addition, in a VC, vehicles can be either the service providers to enrich existing cloud services by providing various on-road information or be the service consumers to enjoy existing centralized cloud services. Therefore, a user can acquire cloud services from either the centralized cloud or the distributed VCs. In papers \cite{Tao_2017_IEEN,Bitam_2015_IWC}, the proposed architecture consists of two hierarchies, permanent cloud, e.g., a powerful and stationary server, and temporary clouds, e.g., vehicles and drivers' devices. In the permanent cloud, three different types of services (i.e., infrastructure-as-a-service (IaaS), platform-as-a-service (PaaS), and software-as-a-service (SaaS)) are provided for CVs. The permanent cloud has powerful and stationary computing and storage capacities and can provide computing, storage, and network resources to the CV system entities. While in temporary clouds, CVs, vehicles' on-board modules, and the drivers' devices are temporarily integrated together to expand the computing capacity. However, this architecture does not meet some requirements such as scalable, reliable, and secure in a large-scale deployment, especially considering vehicle mobility and dynamic participation of mobile computing resources.

\begin{figure}[tp!]
\centering
\includegraphics[width=0.48\textwidth]{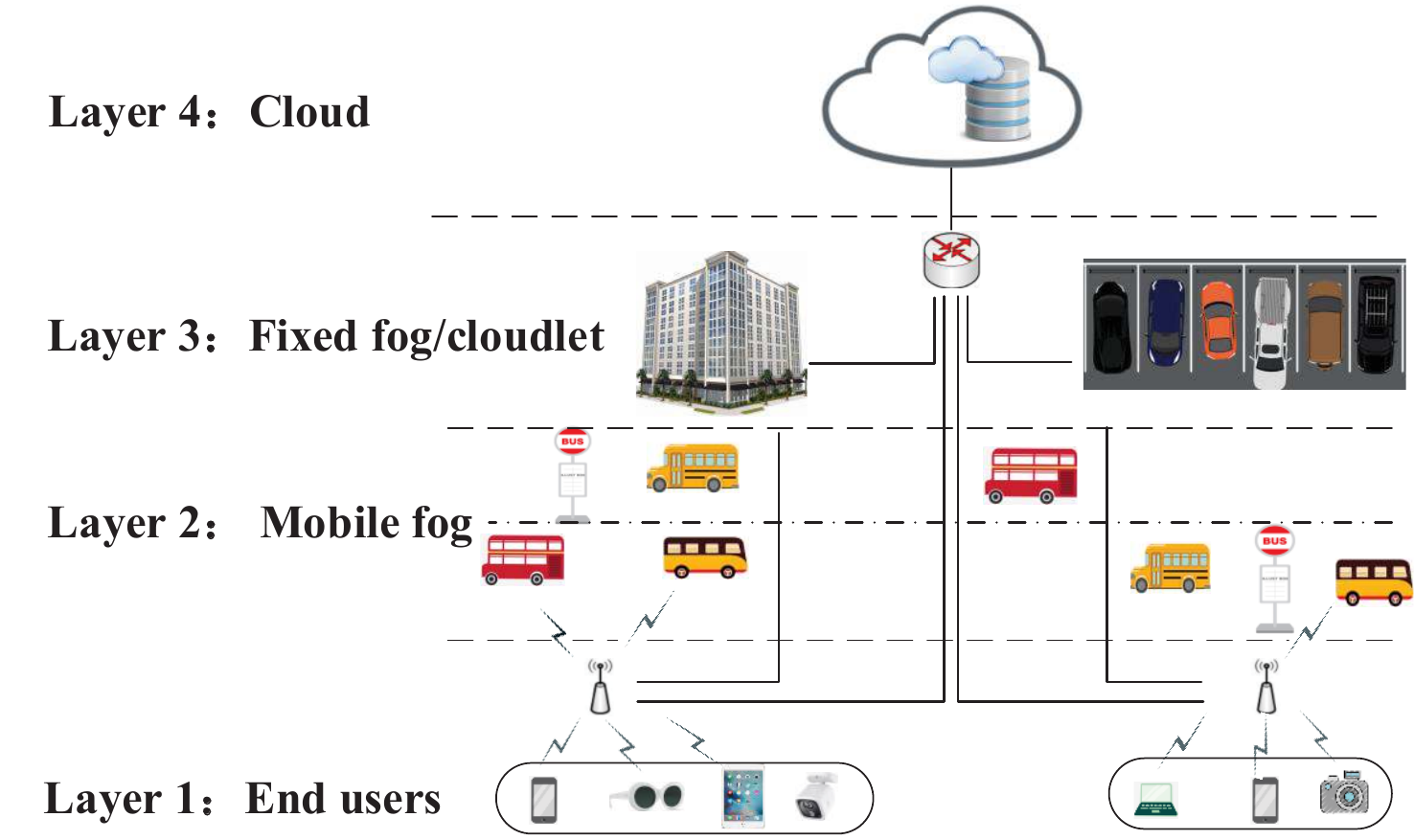}
\caption{The proposed multi-layer computation offloading architecture in paper \cite{wang2018computation}}
\label{Figure8}
\end{figure}

In addition, paper \cite{wang2018computation} proposed a four-layer hybrid architecture which consists of end users, mobile buses, public infrastructures, and remote cloud, as shown in Fig. \ref{Figure8}. Each layer differentiates each other from communication capacity, computation capability, and their inherent properties (e.g., buses have mobility). There is an AP near each bus station, and it serves as the gateway from the users to different kinds of computing resources via the corresponding network interfaces. When a task arrives, users will transmit the task directly to the nearby AP via the wireless network. Then, the collected tasks by the AP can be further offloaded to the mobile fog (i.e., buses in the figure), the fixed fog/cloudlet, or the cloud. The user layer includes people who wear intelligent devices such as smartphones or intelligent glasses, as well as sensor devices such as roadside cameras. They will generate tasks and may not have enough computation resources to execute these tasks. End users may connect to the APs near the bus stations using different network access technologies. The mobile fog layer is formed by a large number of mobile buses that are routinely driving in the city. When buses stop at a station or pass through a station, they will connect to the APs using the millimeter wave technology, offloading tasks from APs. Since these mobile fog nodes are usually located close to end users, they can achieve low latency for uploading and downloading tasks. Once a bus completes the computation tasks, it will return the results directly, if the AP is still in the effective transmission range; otherwise, it will transmit the results to the next AP. The fixed fog layer refers to the infrastructures, such as smart buildings, parking lots, or BSs which usually have fixed locations. Compared with the mobile fog layer, the fixed fog layer may have higher computation capability but longer communication latency. The fixed fog nodes can connect with the mobile fog through the WLAN created by the APs. The cloud layer refers to the cloud data center. It can be physically located in remote areas far from the users. It is usually equipped with powerful computation units but with a high cost of access delay. In traditional cloud applications, users can offload computation tasks to the cloud data center via the wide-area network (WAN).
The characteristics of different layers are summarized in Table \ref{Table11}.

\begin{table*}[tp!]
  \caption{Characteristics of Different Layers in the Proposed Architecture}
  \label{Table11}
  \newcommand{\tabincell}[2]{\begin{tabular}{@{}#1@{}}#2\end{tabular}}
  \centering
\begin{tabular}{|c|c|c|c|}
\hline
Layers \& Property & Communication Capacity & Computation Capability & Availability to Users\\
\hline
Buses & \tabincell{c}{lowest delay, (mmwave, \\3 Gb/s \cite{Baykas_2011_CM})} & \tabincell{c}{Fixed delay, depends on the \\driving time between bus stations\\ (Normal distribution \cite{Liu_2013_TRC})} & \tabincell{c}{Very short distance, \\intermittent, fixed routes}\\ \hline
Vehicles & \tabincell{c}{low delay (2G/3G, mobile wireless \\broadband, WiMAX/WiBro, ZigBee, \\satellite, DSRC, etc.\cite{Dolui_2017_GIoTS})} & \tabincell{c}{Weak, depends on CPU frequency\\ of onboard computer \cite{Zhang_2015_TVT})} & \tabincell{c}{Very short distance, more accessible,\\ randomly driving}\\
\hline
Fixed fog & \tabincell{c}{Medium (WLAN, MAN, \\3G/4G/LTE etc.) \cite{Zhang_2015_TVT}} & Medium & Medium distance\\ \hline
Cloud & \tabincell{c}{Longest delay (WAN, one way Internet \\transmission delay 75-200ms \cite{Wu_2016_TCSVT, Satyanarayanan_2009_PC})} & Unlimited resource & Remote, large number of hops\\ \hline
\end{tabular}
\end{table*}

As shown in Fig. \ref{Figure8}, the concerned task scheduling period starts when there are buses approaching a bus stop. In case there is no bus or vehicles approaching the stop, the tasks will be processed by the fixed fog or the cloud. According to IEEE 802.11ad standard, time division multiple access (TDMA) can be adopted for mmwave transmission among devices \cite{Niu_2015_WN}. When two transmitters (i.e., buses) are close at the bus stop, the corresponding MAC protocol of the adopted wireless access network can manage the interference by coordinating the transmission among different devices towards two transmitters \cite{Qiao_2015_CM}. In addition, highly directional antennas can greatly decrease the interference of concurrent transmissions among mmwave transmitters. Therefore, the interference caused by two buses transmitting simultaneously can be managed. Note that the number of the allocated slots for a task determines the transmission rate of the task. When a bus is approaching a bus station, the achievable data rate increases rapidly.

Security issues in hybrid computation-enabled architectures include security challenges in both computation-aided and distributed computation-enabled architectures, such as trustworthiness between the VC members, misbehaving vehicles, vehicular privacy, etc. Although some of these security challenges are investigated in VANETs and VCs, there are still no solutions that treat all these challenges. For example, in a hybrid computation-enabled architecture, a vehicle can interact with a lot of entities, such as neighboring vehicles, RSUs, conventional cloud/edge server, etc. The vehicle therefore face more danger from the data stealing, hostile attack, and virus infection. Thus, it is crucial to develop a suite of elaborate and comprehensive security solutions for hybrid computation-enabled architectures.

\subsection{Comparison among Computing Architectural Alternatives}
\label{ssc:Com_Cen_Dis}

In this section, we compare the pros and cons of centralized, distributed, and hybrid computing architectures in terms of supporting CV applications, which are summarized in Tables \ref{Table12} and \ref{Table13}, based on the five functional requirements of CV systems explained in Section \ref{sec:Functions}.

\begin{table*}[tp!]
\newcommand{\tabincell}[2]{\begin{tabular}{@{}#1@{}}#2\end{tabular}}
\centering
\caption{The Comparison of Advantages in Centralized, Distributed, and Hybrid Computing Architectural Alternatives for CVs}
\begin{tabular}{|c|c||l|l|l|l|l|}
\hline
\textbf{Sub} & \multirow{2}{*}{\textbf{Category}} & \multicolumn{5}{c|}{\textbf{Advantages}} \\\cline{3-7}
\textbf{Category} &  &  \textbf{Data Sharing} & \textbf{Data Processing} & \textbf{Monitoring} & \textbf{Warning} & \textbf{Control}\\\hline

\multirow{5}{*}{Centralized} & \multirow{5}{*}{\tabincell{c}{Computation-\\aided}} &\tabincell{l}{Comprehensive data transmission \\ approaches (i.e., V2I, V2V, V2D, V2C)} & \tabincell{l}{Unlimited computing power\\ accessible on demand} & \multicolumn{3}{l|}{\tabincell{l}{Large monitoring coverage for data\\ aggregation}} \\\cline{3-7}
& & Stable network connections & \tabincell{l}{Accessible from anywhere} &  \multicolumn{3}{l|}{\tabincell{l}{Large storage space for storing and\\ analysing historic information}} \\\cline{3-7}
& & Secure & \tabincell{l}{No need for early planning\\ of resource provisioning} & \multicolumn{3}{l|}{\tabincell{l}{Secure warning/control (e.g., road\\ condition)}} \\\hline

\multirow{3}{*}{Distributed} & \multirow{3}{*}{\tabincell{c}{Computation-\\enabled}} & High bandwidth & \multirow{3}{*}{Greenness} & \multicolumn{3}{l|}{\tabincell{l}{Fast (i.e., emergent monitoring/warning,\\ e.g., driver's health)}} \\\cline{3-3}\cline{5-7}
&  & Improved network efficiency &  & \multicolumn{3}{l|}{\tabincell{l}{Flexible (i.e., support user-self-defined\\ monitoring/warning/control services)}} \\\hline
 
\multirow{9}{*}{Hybrid} & \multirow{3}{*}{\tabincell{c}{Computation-\\aided}} &  \multirow{3}{*}{\tabincell{l}{Comprehensive data transmission \\ approaches (i.e., V2I, V2V, V2D, V2C)}} & Resilience & \multicolumn{3}{l|}{\tabincell{l}{Heterogeneous data gathering approaches}} \\\cline{4-7}
 &  &  & Low response latency &\multicolumn{3}{l|}{\multirow{2}{*}{\tabincell{l}{Robust}}} \\\cline{4-4}
 &  &  & Improved service agility & \multicolumn{3}{l|}{\tabincell{l}{}} \\\cline{2-7}
 
 & \multirow{5}{*}{\tabincell{c}{Computation-\\enabled}} & \tabincell{l}{Comprehensive data transmission \\ approaches (i.e., V2I, V2V, V2D, V2C)} & Resilience &  \multicolumn{3}{l|}{\tabincell{l}{Heterogeneous data gathering approaches}} \\\cline{3-7}
 &  & \multirow{4}{*}{Information filtration} & Greenness &  \multicolumn{3}{l|}{\tabincell{l}{Robust}} \\\cline{4-7}
 &  &  & Low response latency &  \multicolumn{3}{l|}{\multirow{2}{*}{\tabincell{l}{Flexible}}} \\\cline{4-4}
 &  &  & Improved service agility & \multicolumn{3}{l|}{} \\\hline
\end{tabular}
\label{Table12}
\end{table*}

\begin{table*}[tp!]
\newcommand{\tabincell}[2]{\begin{tabular}{@{}#1@{}}#2\end{tabular}}
\centering
\caption{The Comparison of Disadvantages \& Challenges in Centralized, Distributed, and Hybrid Computing Architectural Alternatives for CVs}
\begin{tabular}{|c|c||l|l|l|l|l|}
\hline
\textbf{Sub} & \multirow{2}{*}{\textbf{Category}} & \multicolumn{5}{c|}{\textbf{Disadvantages \& Challenges}} \\\cline{3-7}
\textbf{Category} &  & \textbf{Data Sharing} & \textbf{Data Processing} & \textbf{Monitoring} & \textbf{Warning} & \textbf{Control}\\\hline

\multirow{2}{*}{Centralized} & \multirow{2}{*}{\tabincell{c}{Computation-\\aided}} & Traffic congestion & \multirow{2}{*}{Long response latency} & \multicolumn{3}{l|}{\multirow{2}{*}{Long response latency}}\\\cline{3-3}
 &  & Constrained bandwidth &  &  \multicolumn{3}{l|}{} \\\hline

\multirow{5}{*}{Distributed} & \multirow{5}{*}{\tabincell{c}{Computation-\\enabled}} &  \tabincell{l}{Limited data transmission\\ approaches (i.e., V2V, V2D)} & Variable computation resource density &   \multicolumn{3}{l|}{\tabincell{l}{Constrained information gathering\\ sources}} \\\cline{3-7}
 &  & Instability & Unreliability \& Heterogeneity & \multicolumn{3}{l|}{\tabincell{l}{Constrained storage space for storing\\ and analyzing historic data}} \\\cline{3-7}
 &  & \tabincell{l}{Insecure (e.g., authentication,\\confidentiality)} & {\tabincell{l}{Frequent and fast resource management\\ required (e.g., frequent scanning)}} &  \multicolumn{3}{l|}{Insecure} \\\hline
 
\multirow{10}{*}{Hybrid} & \multirow{5}{*}{\tabincell{c}{Computation-\\aided}} & Lack of mobility management & \tabincell{l}{High deployment cost of heterogeneous\\ computing infrastructures} &  \multicolumn{3}{l|}{} \\\cline{3-4}
&  & \tabincell{l}{Lack of policy and operational\\ management} & \multirow{3}{*}{\tabincell{l}{Interoperability and standardization\\ required}} & \multicolumn{3}{l|}{} \\\cline{3-3}
&  & \tabincell{l}{Insecure (e.g., identity spoofing,\\information disclosure)} &  & \multicolumn{3}{l|}{} \\\cline{2-7}

& \multirow{5}{*}{\tabincell{c}{Computation-\\enabled}} & Lack of mobility management & \tabincell{l}{High deployment cost of heterogeneous\\ computing infrastructures} &   \multicolumn{3}{l|}{\tabincell{l}{}} \\\cline{3-4}
&  & \tabincell{l}{Lack of policy and operational\\ management} & \tabincell{l}{Interoperability and standardization\\ required} &  \multicolumn{3}{l|}{} \\\cline{3-4}
&  & \tabincell{l}{Insecure (e.g., identity spoofing,\\information disclosure)} & \tabincell{l}{Comprehensive task scheduler required} &  \multicolumn{3}{l|}{} \\\hline
\end{tabular}
\label{Table13}
\end{table*}

\subsubsection{Data Sharing}
\label{ssc:datasharing}
\begin{enumerate}
\item[*] \textbf{Centralized:} \emph{Advantages:} (i) Under centralized computing architectures, V2I, V2V, V2D, and V2C data sharing are all available for CVs. For example, in paper \cite{Wang_2017_TVT}, a metropolitan air quality monitoring service is proposed, where CVs act as the air quality data gathering sensors. Vehicles offload their gathered data to a centralized cloud for performing an air quality estimation. (ii) In addition, centralized architectures provide relatively stable network connections, where CVs can obtain cloud services from anywhere in the world at any time. (iii) Furthermore, centralized computing architectures usually offer greater security over decentralized systems because all of the user data is stored in a central location.

\emph{Disadvantages \& Challenges:} In centralized computing architectures, all of the shared data is directly transmitted to the remote cloud server or MEC servers without any pre-processing or filtering at local. In spite of potential advances in wireless communication technologies (e.g., 5G), ``the bandwidth required for efficient transmission of such a big volume of data is not guaranteed due to a wide range of logistical, political, and geographical factors" \cite{huang2017vehicular}. For example, ``assuming 20 GB per month per vehicle and three million vehicles ($12\%$ market share and $25\%$ regional ratio of 100 million vehicles), 60 petabytes of vehicle data will come to the cloud every month. Assuming the data transaction rate at the cloud is 10 GB per second, it will take 70 days just for the transactions" \cite{AECC_2018_Whitepaper}. Therefore, the constrained bandwidth will be a bottleneck for most vehicular services/applications that transmit a large amount of data or require frequent data sharing.

\item[*] \textbf{Distributed:} \emph{Advantages:} Under distributed computing architectures, direct wireless communications from vehicle to vehicle allow to data to be exchanged fast even where there is no communication infrastructures such as BSs of cellular networks or APs of WLANs. Defined specifically to VANETs, the DSRC that operates in the $5.9$ GHz band is intended to provide high-speed and secure wireless communication for V2V and V2I communications. The U.S. Federal Communications Commission (FCC) allocated $75$ MHz bandwidth at $5.850$-$5.925$ GHz spectrum band for DSRC, while ETSI allocated $70$ MHz in the $5.855$-$5.925$ GHz band. The DSRC can support a CV with a speed up to $200$ km/h, covering a range of $300$ m and reaching up to $1000$ m, and the default data rate is up to $27$ Mbps \cite{cunha2016data,lu2014connected}. IEEE 802.11p wireless access in vehicular environments (WAVE) is the specification of DSRC \cite{jiang2008ieee}. In addition, there have been extensive research work to explore communication properties of DSRC \cite{yin2004performance,taliwal2004empirical,bai2010toward,wu2013vehicular}.

\emph{Disadvantages \& Challenges:} (i) Under distributed computing architectures, CVs only possess very limited data transmission approaches, such as V2V and V2D. Furthermore, the small effective network diameter of VANET leads to a weak connectivity in the communication between CVs. These significantly constrain the data sharing efficiency in many situations. For example, if a CV, named ``red", in vehicle cluster A intends to share its information with another CV, named ``yellow", in vehicle cluster B, its data can only be successfully transmitted to ``yellow" through multi-hop V2V connections. Furthermore, in urban areas, the line-of-sight (LOS) path of V2V may often be blocked by buildings at intersections. (ii) One of the major challenges of distributed computing architectures for CVs is the instability of communication connections among CVs. Because of the high mobility, there is no guarantee on the behaviors of the vehicle and the highly dynamic topology results in frequent changes in its connectivity. Therefore, the connection between two CVs can be quickly interrupted when they are transmitting data. Additionally, high vehicle mobility also leads to Doppler effects, which may cause severe wireless loss. (iii) In addition, there are several challenges that threaten CVs' security under distributed computing architectures. Providing secure network connections in distributed architectures is more difficult than in centralized architectures because of the high mobility of CVs \cite{karagiannis2011vehicular}. For example, in VANET-based distributed architectures, the security issues can be classified into five categories \cite{whaiduzzaman_2014}: authentication \cite{kamat2006identity,calandriello2007efficient,verma2009segcom}, non-repudiation \cite{intelligent2006ieee,raya2006efficient}, confidentiality \cite{huang2010situation}, verification of data \cite{golle2004detecting}, and localization \cite{harsch2007secure, vora2006secure}.

\item[*] \textbf{Hybrid (Computation-aided):} \emph{Advantages:} Hybrid architectures also offer comprehensive data sharing approaches including V2I, V2V, V2D, and V2C.

\emph{Disadvantages \& Challenges:} (i) One of the major characteristics of VCs is the high mobility, which may cause several challenges in different fields including routing protocol designing, security, and data transmission reliability, especially in hybrid architectures. (ii) Due to the large diversity of networks in the hybrid computing architecture (e.g., mobile ad hoc networks, wireless sensor networks, VANETs, etc.), comprehensive policy and operational management should be established. For instance, we are expecting that several types of networks will emerge and they will interact with each other seamlessly. (iii) Security issues in hybrid CV systems are more complicated than in other networks (e.g., wireless sensor networks) due to the high mobility of CVs and the large diversity of networks. For example, it is difficult to verify the integrity of messages and authentication of users when vehicles move fast \cite{whaiduzzaman_2014}. Therefore, security issues in both network (e.g., VANET) and transmission (e.g., wireless communication channel) layers in hybrid computing architectures need more consideration.

\item[*] \textbf{Hybrid (Computation-enabled):} \emph{Advantages:} Localized computation resources such as fog nodes and on-board computing units can pre-process the collected data of CVs. Thus, the collected data can be aggregated/processed/filtered before uploading to the cloud, which reduces the offloaded data volume and meanwhile improves the efficiency of the network.

\emph{Disadvantages \& Challenges:} (i), (ii), and (iii) presented in Hybrid (Computation-aided).
\end{enumerate}

\subsubsection{Data Processing}

\begin{enumerate}
\item[*] \textbf{Centralized:} \emph{Advantages:}
(i) The remote cloud server and MEC servers under centralized computing architectures possess large storage space and huge computation power. Therefore, centralized computing architectures can provide CVs the facility of unlimited computing power accessible on demand, which is suitable for vehicle applications that need large storage capacity and computation power. (ii) Under centralized architectures, CVs are capable of accessing servers' computation resources from anywhere at any time. Because of the high mobility characteristic of vehicles, this benefit is significantly important for CVs to obtain stable and uninterrupted services. (iii) Furthermore, centralized architectures do not require early planning of computation resource provisioning.

\emph{Disadvantages \& Challenges:} As we mentioned previously, the current trend of concentrating data processing at centralized cloud servers will cause huge data transmission traffic. This will directly incur unnecessarily long response delay and in turn, will increase the computation latency \cite{AECC_2018_Whitepaper}. Therefore, CV applications, such as intelligent driving and high resolution map creation and distribution \cite{AECC_2018_Whitepaper}, which are latency-sensitive cannot be handled effectively under centralized computing architectures.

\item[*] \textbf{Distributed:} \emph{Advantages:} In centralized architectures, the cloud data server consumes a huge amount of energy each year \cite{zhang2010cloud}. In contrast, distributed architectures enable green computing by efficiently using the spare on-board computation resources on CVs. Since vehicles are self-powered, efficiently using on-board computation resources on CVs will help minimize the energy consumption of cloud/edge servers.

\emph{Disadvantages \& Challenges:}
(i) The computation resource density in distributed architectures varies depending on the traffic density, which can be very high in the case of a traffic jam, or very low, as driving on the highway. However, when a vehicle stops or moves slowly, it does not require more computation power (e.g., a CV does not need to rapidly process or refresh its high-definition map for supporting intelligent driving). In contrast, when the vehicle is moving fast, it requires much computation power to accurately, frequently, and rapidly process its sensors captured data (e.g., accurately localize its surrounding dynamic objects). (ii) A VC has the issue that its computation resource availability is not reliable. Since the computation resource availability in a VC and the behavior of each CV are not guaranteed. For instance, vehicles might unexpectedly leave or join a VC. In fact, this is also one of the major differences between the distributed architecture and the centralized architecture. In addition, different CVs in a VC usually have different characteristics or capabilities (e.g., processor speed and memory volume) upon their manufacturers, models, and applications. For example, CVs can be classified into private vehicles and buses. Private vehicles have unpredictable routes, whereas buses have fixed routes. (iii) Because of the variable computation resource density and the unreliability of the computation resources in distributed architectures, it is critical for every CV to frequently monitor its cooperated CVs and to periodically search for potential candidates.

\item[*] \textbf{Hybrid (Computation-aided):} \emph{Advantages:} (i) Compared to the centralized and distributed architectures, the hybrid architecture provides CVs with more potential computation sources, which improves the resilience of CV systems. (ii) The hybrid architecture also can provide low response latency for CVs that require running latency-sensitive applications. Instead of offloading collected data to a remote cloud, CVs can choose to process their data at neighboring fog nodes or RSUs.

\emph{Disadvantages \& Challenges:} (i) High deployment cost of heterogeneous computing infrastructures. (ii) Interoperability and standardization required. Since the hybrid architecture is based on diverse stationary and mobile computing resources, many steps should be taken to address the interoperability challenge to allow these different entities to work together. In addition, standardization is a potential solution to address the interoperability issue so that a consensus can be reached among developers, vehicle manufacturers, etc.

\item[*] \textbf{Hybrid (Computation-enabled):} \emph{Advantages:} (i), (ii), and (iii) presented in Hybrid (Computation-aided) and greenness presented in Distributed.

\emph{Disadvantages \& Challenges:} (i) and (ii) presented in Hybrid (Computation-aided). (iii) A comprehensive task scheduler is required in hybrid computing architectures. As we presented above, a variety of computing units/sources coexist in a hybrid CV system, which reflects the situation in which a task generated by a CV can be processed by different approaches. Such a situation will lead to an issue that what is the best computing approach for each task or what are the criteria of the processing decision making. From the perspective of the CV application, for instance, it may depend on the complexity of the application, requirement of the response latency, transmitted data volume, required computation capacity, energy consumption, monetary cost, etc. While, from the perspective of the system, it may depend on the position and role of the CV in the hierarchical system deployment, load of each computing sources, accessibility of each computing units, etc. Therefore, designing a comprehensive task scheduler that can cover multiple criteria from different perspectives is essential and challenging in hybrid architectures.
\end{enumerate}

\subsubsection{Monitoring, Warning, and Control}
\begin{enumerate}
\item[*] \textbf{Centralized:} \emph{Advantages:} (i) Since all vehicles under centralized computing architectures are covered by a cloud server or several MEC servers, it is efficient to monitor the presence and experience of CVs under this architecture. For example, global warning services, such as warnings of traffic jam, accident, road condition, and predicted weather event, can be provided by the remote cloud server. (ii) In addition, by using the centralized cloud storage that can gather a huge amount of user historic information, various governmental and private agencies can use the gathered data to provide various monitoring and warning services or perform diverse studies. Furthermore, control services that need historical information of CVs can be implemented under this architecture. For example, the centralized server can make the control decision on whether a vehicle should enter the automated driving mode or not, based on not only the current collected data about the vehicle, such as the speed, but also the historical data, such as road conditions.

\emph{Disadvantages \& Challenges:} since the centralized servers are located relatively far away from the CVs, the delay of receiving control messages is relatively long, which means that centralized computing architectures may not be suitable for delay-sensitive services, such as autonomous driving.

\item[*] \textbf{Distributed:} \emph{Advantages:} (i) Comparing to the centralized architecture, the distributed architecture, such as VC, has great advantage in emergency warning scenarios. Disaster management, for instance, is an emergent warning application, where CVs send or broadcast the warning messages to all resources such as nearby vehicles and authorized authorities in case of any disaster. (ii) Distributed architectures provide flexibility for CVs to design user-self-defined monitoring/warning/control services. For example, it is easier for a user to set up a driver's health monitoring service with its own concerns, such as driver's medical records and habits. In addition, this monitoring service can still work even without an Internet connection.

\emph{Disadvantages \& Challenges:} (i) Unlike the previously described global warning services under centralized computing architectures, distributed computing architectures can only provide one type of local warning services, driver health warning, because of the constraint of VC coverage. (ii) The limited storage space constrains the  ability of storing and analyzing a large amount of user historic data.

\end{enumerate}

\begin{figure*}[t]
\centering
\subfigure[] 
{\includegraphics[width=0.46\textwidth]{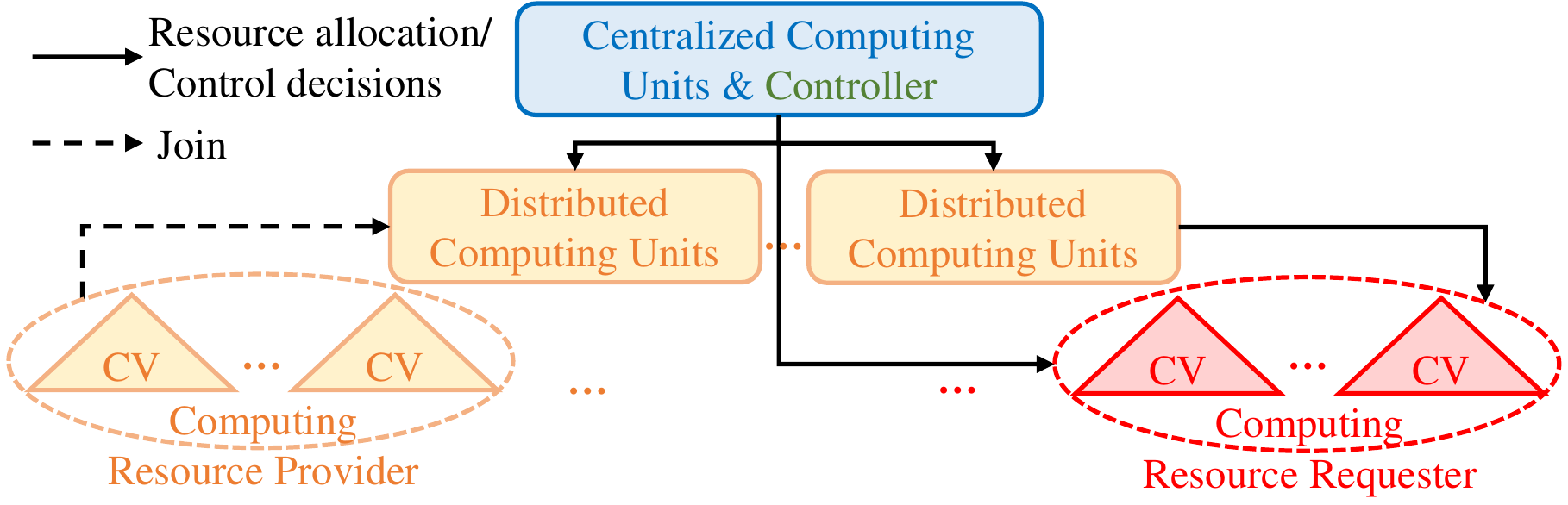}\label{Figure9(1)}} 
\subfigure[]
{\includegraphics[width=0.44\textwidth]{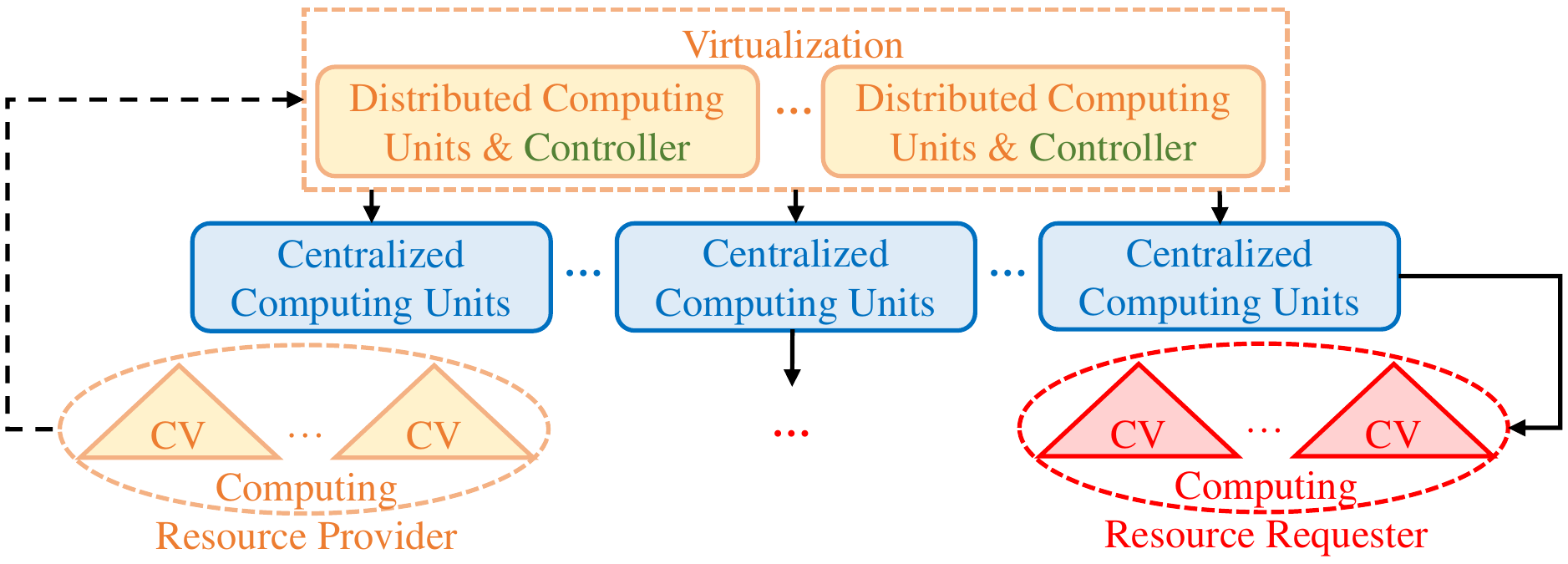}\label{Figure9(2)}}
\caption{Other hybrid architectural alternatives.}
\label{Figure9}
\end{figure*}

\section{Summary and Open Research Issues}\label{sec:openissue}

\subsection{Summary}
\begin{itemize}
\item The effective design of computing architectures for CVs should integrate advanced techniques from both areas of wireless communications (e.g., 5G, V2V, and V2I) and mobile computing (e.g., cloud/edge/fog computing).
\item It is critical to choose suitable computing architectures for different CV applications or services. For example, the centralized computation-aided computing architecture can be applied for intelligent transportation systems which require complicated traffic management strategies and massive transport data but is not suitable for autonomous driving services due to the stringent computation latency requirements.
\item Comparing to computation-aided architectures, computation-enabled architectures have the potential to provide more flexible and lower energy/monetary cost computation services for CVs due to the variant computation resource providers and short distances between the computation resource requesters and providers. However, they are still facing a lot of challenges, such as fast mobility support, stability (i.e., dynamic participation of mobile computing resources), and security. 
\end{itemize}

\subsection{Open Research Issues}

Designing an appropriate computing architecture for CVs is one key research to successfully provide a wide range of CV services by meeting their QoS requirements.
Hence, we believe that existing architectural alternatives introduced in this paper can be a good starting point to build a future CV eco-system that utilizes a large amount of data generated by vehicles. However, designing a good computing architecture cannot entirely solve all the issues that we will encounter in the future. We conclude this paper by listing several open challenges and research directions.

\subsubsection{Data sharing}
\textbf{Localizing data traffic:}
In order for OEMs to provide diverse and personalized connected services, it is essential to collect various types of sensor data from each vehicle such as camera images, acceleration/deceleration or hard braking.
Those services may include personalized insurance service, predictive vehicle maintenance, high-definition (HD) map generation, intelligent driving and so on. Some forecasts that each vehicle will generate data in the order of gigabytes every day. If those data are directly sent to cloud servers, it will add a huge amount of network traffic in the existing infrastructure.
Even though a vehicle can produce the raw data at a high frequency and volume, it may not be necessary to forward every single data. For example, some data are only meaningful in a particular geographic region (e.g., a hard braking signal to avoid frontal collision in the highway); it may be also fine to send data less frequently to the server (e.g., analyzing drivers’ behavior for a personalized insurance service). One research direction is to precisely characterize various types of connected services to determine the sufficient amount of frequency and volume of data enough to achieve service-specific QoS. Such knowledge can then be distributed across the multiple layers of the system architecture to appropriately localize data to avoid excessive and unnecessary traffic in the overall network.

\textbf{Highly heterogeneous vehicular networks:} Due to the diverse QoS requirements of vehicular services and unique characteristics of vehicular network connections, such as high mobility, frequent topology changes, and unreliable connectivity, homogeneous radio access technology (e.g., either DSRC or LTE) cannot satisfy the performance requirements of vehicular services. For example, DSRC is capable of providing low round-trip time (RTT) (i.e., below 10 milliseconds \cite{xu2017dsrc}), which satisfies the latency requirement of safety-related vehicular services. However, its limited peak data rate and signal coverage decline the efficiency and reliability of data sharing among CVs and infrastructures. Furthermore, although cellular networks such as LTE offers a much wider signal coverage and a higher peak data rate for CVs compared to DSRC, stringent latency requirement cannot be guaranteed due to its long RTT (i.e., over 300 milliseconds \cite{xu2017dsrc}). Therefore, in order to meet different QoS requirements of vehicular services, future vehicular networks tend to be highly heterogeneous. However, several open issues are imposed by the network heterogeneity. (i) Developing radio access method selection, link adaptation, and radio resource management schemes in heterogeneous vehicular networks are crucial and complicated. Thus, the main challenge is how to balance performance and complexity. (ii) Maintaining a seamless connectivity across diverse radio access technologies is complex, which includes inter- and intra-radio handoffs. Most traditional handoff approaches for cellular networks are centralized and provide a single trigger mechanism, which is not sufficient to support 
distributed and hybrid vehicular environments. 

\textbf{Support mobility in heterogeneous architectures:}
Mobility is one key feature of the automotive system; a vehicle continuously moves around different regions expecting to receive services seamlessly. However, it is practically difficult to expect the vehicle to communicate with a uniform infrastructure or architecture everywhere for several practical reasons. For example, different organizations (e.g., government, network providers, OEMs) may want to own their infrastructures exclusively due to security concerns; an infrastructure may be installed in a limited area only due to a budget issue or lack of profitability. Such situations will cause a vehicle to communicate with very different infrastructure as it moves.
Another research direction is to develop a way to provide seamless CV services across heterogeneous infrastructure or architectures. This requires each infrastructure to provide the open interface that specifies which types of services can be supported or prohibited more explicitly, instead of hiding such information. Then, each CV service provider may have a better understanding as to where the infrastructure support for the service is available, and if it is not available, what are the alternative way to provide the service, such as service hand-over and migration. Making such interfaces available across different organizations will also boost the adoption of CVs.

\subsubsection{Data processing}
\textbf{Computing resource management:}
Each individual subsystem (e.g., cloud servers, fog/edge servers, RSUs, and vehicles) in the surveyed architectures has a different limitation of the computing resources such as memory or computation power.
On the other hand, each connected service may require different types and sizes of computing resources.
For example, the over-the-air (OTA) update service may require fog servers to reserve a portion of memory storage that can be used to distribute the latest version of the software patch to the vehicles in a particular area; the HD map generation requires relatively a higher computation power to synthesize local maps at real-time.
Guaranteeing QoS of various services should take into account such computing resource constraints imposed on each subsystem.
The relevant research direction is to manage computing resources in a more rigorous way, and the example research may include the following. If a particular area experiences a relatively higher request on a certain service, one can allocate more resources on the infrastructure in the area (resource allocation); depending on the current workload on the infrastructure, one can decide to (or not to) offload a certain computation (data/computation offloading); one can predict the future usage of the resources by reading the past trend of the requested connected services (resource provisioning). If this type of research is combined with the surveyed architectures, we expect a better quality of CV services can be provided.

\subsubsection{Monitoring, warning, and control}

\textbf{Other hybrid architectural alternatives:} As we discussed in Section \ref{sec:Lit_Survey}, architectural design is significantly important for acquiring diverse system features and satisfying different service requirements. Although a large number of architectural alternatives for CV systems have been proposed in existing work, there should be more possibilities, especially for the hybrid architectural design. For example, as shown in Fig. \ref{Figure9(1)}, most of existing hybrid architectures choose to put the centralized computing in the highest layer, which acts as a remote cloud. However, besides this stereotypical architectural design, we may have another alternative, as shown in Fig. \ref{Figure9(2)}, where distributed computing is in the highest layer in the CV system. The centralized computing units are located closer to CVs (e.g., RSUs) rather than located in remote cloud servers. Such design will (i) make the CV system fully utilize the computation and storage capacities of centralized computing units; (ii) reduce both communication and computation latency. Furthermore, distributed computing units, such as spare computing resources in neighbouring CVs and fog nodes, can be combined as a virtual localized controller which is responsible for allocating the computing resources in its governed centralized computing units to CVs. Therefore, based on the above discussion, there might be a set of interesting research opportunities in CV system computing architectural design, which are not yet tackled in the reviewed literature. 

\section{Conclusion}
\label{sc:conclusion}
CV has a great potential to provide safer and more comfortable driving experience with the expected service scenarios, such as intelligent driving, V2C cruising, high-definition map generation, etc. In contrast to existing related surveys which only focus on one specific computing architecture in the whole paper and lack discussions on benefits, research challenges, and system requirements of different architectural alternatives, this paper has presented a thorough study on architectural design based on cloud/edge/fog computing for CVs. We have comprehensively surveyed and compared the state-of-the-art architectural alternatives with the goal of understanding the benefits and challenges of each architectural design within different CV applications. However, given the relative infancy of the field, there are still a number of outstanding problems that require further investigation from the perspective of advanced solutions including other hybrid architectural alternatives, localizing data traffic, mobility support in heterogeneous architectures, and computing resource management.

\balance
\bibliographystyle{IEEEtran}
\bibliography{reference}
\end{document}